\documentclass{article}          % twocolumn
\usepackage{authblk}
\usepackage{fullpage}
\usepackage{amsthm}
\usepackage{graphicx}
\usepackage{url}
\usepackage{xcolor}
\usepackage{natbib}
\usepackage{amsmath}
\usepackage{amssymb}
\usepackage{arydshln}
\usepackage{hyperref}
\usepackage{listings}
\usepackage{rotating}

\definecolor{dkgreen}{rgb}{0,0.6,0}
\definecolor{gray}{rgb}{0.5,0.5,0.5}
\definecolor{mauve}{rgb}{0.58,0,0.82}

\newcommand{\code}[1]{\texttt{#1}}
\newcommand{\proglang}[1]{\textsf{#1}}
\newcommand{\pkg}[1]{{\fontseries{b}\selectfont #1}}

\newcommand{\Expect}[1]{\mathbb{E} \left[{#1}\right]}

\newcommand{\Cov}[1]{\mbox{Cov} \left[{#1}\right]}

\newcommand{\Kxx}{\mathbf{K}_{\mathbf{*,*}}}
\newcommand{\Kxf}{\mathbf{K}_{\mathbf{*,f}}}
\newcommand{\Kfx}{\mathbf{K}_{\mathbf{f,*}}}
\newcommand{\Qxf}{\mathbf{Q}_{\mathbf{*,f}}}
\newcommand{\Qfx}{\mathbf{Q}_{\mathbf{f,*}}}
\newcommand{\Kff}{\mathbf{K}_{\mathbf{f,f}}}
\newcommand{\Qxx}{\mathbf{Q}_{\mathbf{*,*}}}

\newcommand{\Qff}{\mathbf{Q}_{\mathbf{f,f}}}
\newcommand{\Kuu}{\mathbf{K}_{\mathbf{u, u}}}
\newcommand{\Kfu}{\mathbf{K}_{\mathbf{f, u}}}

\newcommand{\Kxu}{\mathbf{K}_{\mathbf{*, u}}}

\newcommand{\bx}{\mathbf{x}}
\newcommand{\bI}{\mathbf{I}}
\newcommand{\bff}{\mathbf f}

\newcommand{\bX}{\mathbf X}
\newcommand{\by}{\mathbf y}
\newcommand{\bv}{\mathbf v}
\newcommand{\bu}{\mathbf u}

\newcommand{\given}{\,|\,}
\newcommand{\btheta}{\boldsymbol\theta}

\begin{document}
\title{\pkg{GaussianProcesses.jl}: A Nonparametric Bayes package for the \proglang{Julia} Language}

\author[1]{Jamie Fairbrother}
\author[2]{Christopher Nemeth\footnote{corresponding author: c.nemeth@lancaster.ac.uk}}
\author[3]{Maxime Rischard}
\author[4]{Johanni Brea}
\author[2]{Thomas Pinder}

\affil[1]{Department of Management Science, Lancaster University, Lancaster, UK}
\affil[2]{Department of Mathematics and Statistics, Lancaster University, Lancaster, UK}
\affil[3]{Department of Statistics, Harvard University, Cambridge, USA}
\affil[4]{Laboratory of Computational Neuroscience, EPFL, Lausanne, Switzerland}

\maketitle

%% an abstract and keywords
\begin{abstract}
  Gaussian processes are a class of flexible nonparametric Bayesian tools that are widely used across the sciences, and in industry, to model complex data sources. Key to applying Gaussian process models is the availability of well-developed open source software, which is available in many programming languages. In this paper, we present a tutorial of the GaussianProcesses.jl package that has been developed for the Julia programming language. GaussianProcesses.jl utilises the inherent computational benefits of the Julia language, including multiple dispatch and just-in-time compilation, to produce a fast, flexible and user-friendly Gaussian processes package. The package provides many mean and kernel functions with supporting inference tools to fit exact Gaussian process models, as well as a range of alternative likelihood functions to handle non-Gaussian data (e.g. binary classification models) and sparse approximations for scalable Gaussian processes. The package makes efficient use of existing Julia packages to provide users with a range of optimization and plotting tools.
  
\end{abstract}


\section{Introduction}
\label{sec:introduction}

% Gaussian processes 
Gaussian processes (GPs) are a family of stochastic processes which provide a flexible nonparametric tool for modelling data. In the most basic setting, a Gaussian process models a \emph{latent function} based on a finite set of observations. The Gaussian process can be viewed as an extension of a multivariate Gaussian distribution to an infinite number of dimensions, where any finite combination of dimensions will result in a multivariate Gaussian distribution, which is completely specified by its mean and covariance functions. The choice of mean and covariance function, also known as the \emph{kernel}, impose smoothness assumptions on the latent function of interest and determines the correlation between output observations $\by$ as a function of the Euclidean distance between their respective input data points $\bx$.

Gaussian processes have been widely used across a vast range of scientific and industrial fields, for example, to model astronomical time series \citep{foreman2017fast} and brain networks \citep{doi:10.1093/bioinformatics/btx050}, or for improved soil mapping \citep{gonzalez2007creating} and robotic control \citep{deisenroth2015gaussian}. Arguably, the success of Gaussian processes in these various fields stems from the ease with which scientists and practitioners can apply Gaussian processes to their problems, as well as the general flexibility afforded to GPs for modelling various data forms. 

% GP Packages
% - We should add some details about the capabilities and weakneses of these [JF]
Gaussian processes have a longstanding history in geostatistics \citep{matheron1963principles} for modelling spatial data. However, more recent interest in GPs has stemmed from the machine learning and other scientific communities. In particular, the successful uptake of GPs in other areas has been a result of high-quality and freely available software. There are now a number of excellent Gaussian process packages available in several computing and scientific programming languages. One of the most mature of these is the \pkg{GPML} package \cite{RasmussenNickisch2017} for the \proglang{MATLAB} language which was originally developed to demonstrate the algorithms in the book by \cite{rasmussen} and provides a wide range of functionality. Packages written for other languages, including \proglang{Python} packages, e.g. \pkg{GPy} \cite{gpy2014} and \pkg{GPFlow} \cite{MatthewsEtAl2017}, have incorporated more recent developments in the area of Gaussian processes, most notably implementations of sparse Gaussian processes.

% Julia
This paper presents a new package, \pkg{GaussianProcesses.jl}, for implementing Gaussian processes in the recently developed \proglang{Julia} programming language. \proglang{Julia} \citep{BezansonEtAl2017}, an open source programming language, is designed specifically for numerical computing and has many features which make it attractive for implementing Gaussian processes.
Two of the most useful and unique features of \proglang{Julia} are \emph{just-in-time (JIT) compilation} and \emph{multiple dispatch}.
JIT compilation compiles a function into binary code the first time it is used, which allows code to run much more efficiently compared with interpreted code in other dynamic programming languages. This provides a solution to the ``two-language'' problem: in contrast to e.g. \proglang{R} or \proglang{Python}, where performance-critical parts are often delegated to libraries written in \proglang{C/C++} or \proglang{Fortran}, it is possible to write highly performant code in \proglang{Julia}, while keeping the convenience of a high-level language.
Multiple dispatch allows functions to be dynamically dispatched based on inputted arguments. 
In the context of our package, this allows us to have a general framework for operating on Gaussian processes, while allowing us to implement more efficient functions for the different types of objects which will be used with the process.
Similar to the \proglang{R} language, \proglang{Julia} has an excellent package manager system which allows users to easily install packages from inside the \proglang{Julia} REPL as well as many well-developed packages for statistical analysis \citep{JuliaStats}.

% Package
% - Should we define kernel and likelihoods above before mentioning here? [JF]
\pkg{GaussianProcesses.jl} is an open source package which is entirely written in \proglang{Julia}. It supports a wide choice of mean, kernel and likelihood functions (see Appendix \ref{sec:available-functions}) with a convenient interface for composing existing functions via summation or multiplication. The package leverages other \proglang{Julia} packages to extend its functionality and ensure computational efficiency. For example, hyperparameters of the Gaussian process are optimized using the \pkg{Optim.jl} package \citep{Mogensen2018} which provides a range of efficient and configurable unconstrained optimization algorithms; prior distributions for hyperparameters can be set using the \pkg{Distributions.jl} package \citep{Distributions.jl-2019}. Additionally, this package has now become a dependency of other \proglang{Julia} packages, for example, \pkg{BayesianOptimization.jl}, a demo of which is given in Section \ref{sec:bayes-optim}. The run-time speed of \pkg{GaussianProcesses.jl} has been heavily optimized and is competitive with other rival packages for Gaussian processes. A run-time comparison of the package against \pkg{GPML} and \pkg{GPy} is given in Section \ref{sec:comp-other-pack}.
% Paper stucture
% To be completed [JF]

The paper is organised as follows. Section \ref{sec:gaussian-processes} provides an introduction to Gaussian processes and how they can be applied to model Gaussian and non-Gaussian observational data. Section \ref{sec:package} gives an overview of the main functionality of the package which is presented through a simple application of fitting a Gaussian process to simulated data. This is then followed by  five application demos in Section \ref{sec:illustration} which highlight how Gaussian processes can be applied to classification problems, time series modelling, count data, black-box optimization and computationally-efficient large-scale nonparametric modelling via sparse Gaussian process approximations. Section \ref{sec:comp-other-pack} gives a run-time comparison of the package against popular alternatives which are listed above. Finally, the paper concludes (Section \ref{sec:future-developments}) with a discussion of ongoing package developments which will provide further functionality in future releases of the package. 

\section{Gaussian processes in a nutshell}
\label{sec:gaussian-processes}

Gaussian processes are a class of models which are popular tools for nonlinear regression and classification problems. They have been applied extensively in scientific disciplines ranging from modelling environmental sensor data \citep{Osborne2008a} to astronomical time series data \citep{Wilson2015} all within a Bayesian nonparametric setting. A \textit{Gaussian Process} (GP) defines a distribution over functions, $p(f)$, where $f: \mathcal{X} \rightarrow \mathbb{R}$ is a function mapping from the input space $\mathcal{X}$ to the space of real numbers. The space of functions $f$ can be infinite-dimensional, for example when $\mathcal{X} \subseteq \mathbb{R}^d$, but for any subset of inputs $\mathbf{X}=\{\bx_1,\bx_2,\ldots,\bx_n\} \subset \mathcal{X}$ we define $\bff := \{f(\bx_i)\}_{i=1}^n$ as a random variable whose marginal distribution $p(\bff)$ is a multivariate Gaussian.

The Gaussian process framework provides a flexible structure for modelling a wide range of data types. In this package we consider models of the following general form, 
\begin{eqnarray}
\label{eq:framework}
\mathbf{y} \given \bff, \btheta &\sim& \prod_{i=1}^n p(y_i\,\given \,f_i,\btheta), \nonumber \\
f(\bx) \given \btheta &\sim& \mathcal{GP}\left(m_{\btheta}(\bx), k_{\btheta}(\bx, \bx')\right),\\
      \btheta &\sim& p(\btheta), \nonumber
\end{eqnarray}
where $\by=(y_1,y_2,\ldots,y_n) \in \mathcal{Y}$ and $\bx \in \mathcal{X}$ are the observations and covariates, respectively, and $f_i:=f(\bx_i)$. We assume that the responses $\by$ are independent and identically distributed, and as a result, the likelihood $p(\by \given \bff, \btheta)$ can be factorised over the observations. For the sake of notational convenience, we let $\btheta \in \mathbf{\Theta} \subseteq \mathbb{R}^d$ denote the vector of model parameters for both the likelihood function and Gaussian process prior. 

The Gaussian process prior is completely specified by its mean function $m_{\btheta}(\bx)$ and covariance function $k_{\btheta}(\bx, \bx')$, also known as the \textit{kernel}. The mean function is commonly set to zero (i.e. $m_{\btheta}(\mathbf{x})=0, \forall \mathbf{x}$), which can often by achieved by centring the observations (i.e. $\by - \Expect{\by}$) resulting in a mean of zero. If the observations cannot be re-centred in this way, for example if the observations display a linear or periodic trend, then the zero mean function can still be applied with the trend modelled by the kernel function. 

\begin{figure}[h]
  \centering
  \includegraphics[scale=0.8]{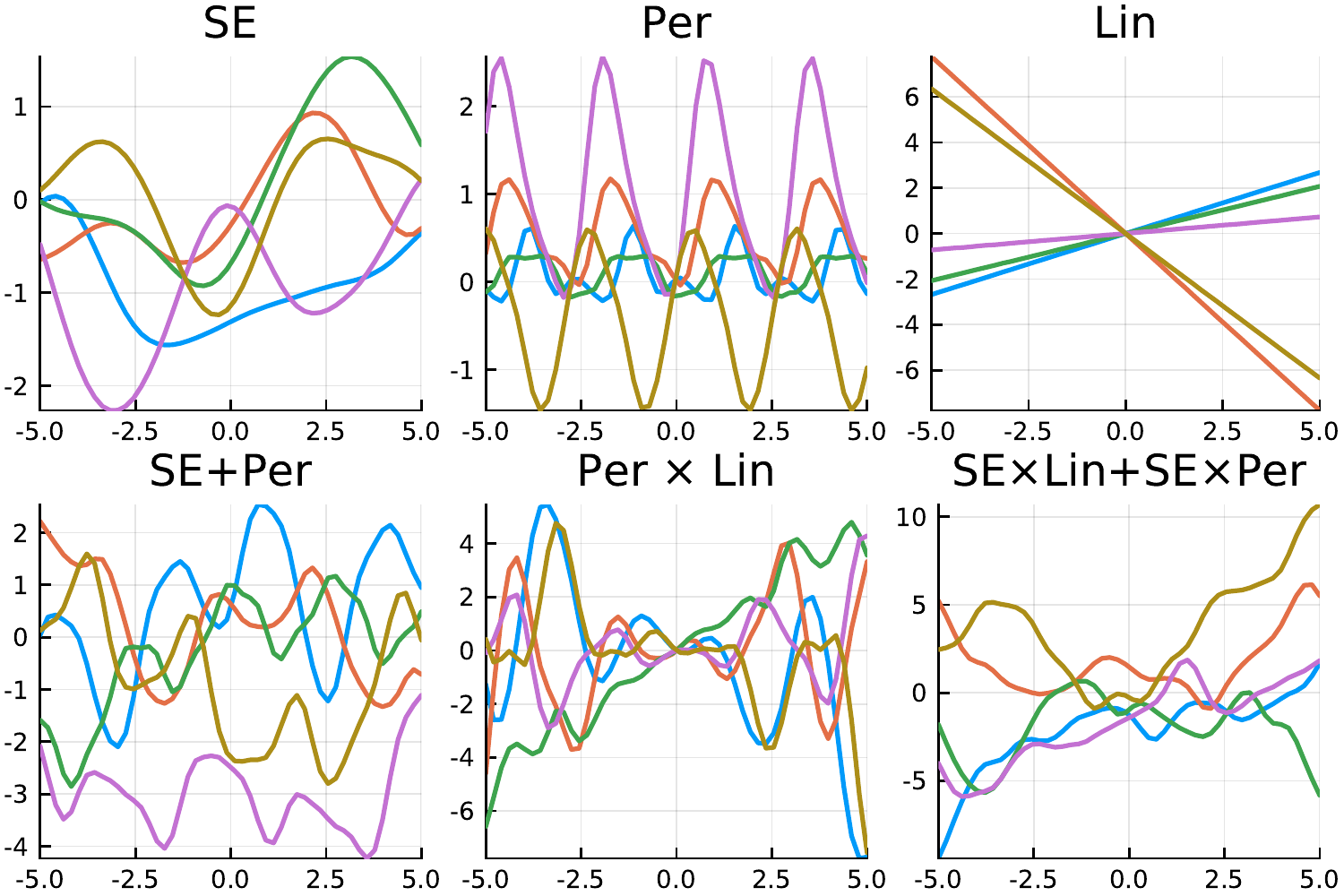}
  \label{fig:kernels}
  \caption{Clockwise from the top left. Five random samples from the Gaussian process prior using the following kernels (refer to the help file for details): \code{SE(0.5,0.0)}, \code{Periodic(0.5,0.0,1.0)}, \code{Lin(0.0)}, \code{SE(0.5,0.0) * Lin(0.0) + SE(0.5,0.0) * Periodic(0.5,0.0,1.0)}, \code{Periodic(0.5,0.0,1.0) * Lin(0.0)} and \code{SE(0.5,0.0) + Periodic(0.5,0.0,1.0)}}
\end{figure}

The kernel determines the correlation between any two function values $f_i$ and $f_j$ in the output space as a function of their corresponding inputs $\mathbf{x}_i$ and $\mathbf{x}_j$. The user is free to choose any appropriate kernel that best models the data as long as the covariance matrix formed by the kernel is symmetric and positive semi-definite. Perhaps the most common kernel function is the \textit{squared exponential}, $\Cov{f(\mathbf{x}),f(\mathbf{x}')} = k(\mathbf{x},\mathbf{x'}) = \sigma^2\exp(-\frac{1}{2\ell^2}|\mathbf{x}-\mathbf{x'}|^2)$. For this kernel the correlation between $f_i$ and $f_j$ is determined by the Euclidean distance between $\mathbf{x}_i$ and $\mathbf{x}_j$ and the hyperparameters $\btheta =(\sigma, \ell)$, where $\ell$ determines the speed at which the correlation between $\mathbf{x}$ and $\mathbf{x'}$ decays. There exists a wide range of kernels that can flexibly model a wide range of data patterns. It is possible to create more complex kernels from the sum and product of simpler kernels \citep{Duvenaud2014}, (see Chapter 4 of \cite{rasmussen} for a detailed discussion of kernels). Figure \ref{fig:kernels} shows one-dimensional Gaussian processes sampled from three simple kernels (squared exponential, periodic and linear) and three composite kernels, and demonstrates how the combination of these kernels can provide a richer correlation structure to capture more intricate function behaviour.

Often we are interested in predicting function vales $\bff^*$ at new inputs $\mathbf{x}^*$. Assuming a finite set of function values $\bff$, the joint distribution between these observed points and the test points $\bff^*$ forms a joint Gaussian distribution,
\begin{equation}
  \label{eq:joint}
  \left( \begin{matrix} \bff \\ \bff^* \end{matrix} \right) \given \mathbf{X}, \mathbf{x}^*,\btheta 
  \sim \mathcal{N}\left(\begin{matrix} 0, \left[\begin{matrix}
          \Kff   &  \Kfx \\
          \Kxf  &  \Kxx) \\
        \end{matrix}\right]
    \end{matrix}\right),
\end{equation}
where $\Kff = k_{\btheta}(\mathbf{X}, \mathbf{X})$, $\Kfx = k_{\btheta}(\mathbf{X}, \mathbf{x}^*)$ and $\Kxx = k_{\btheta}(\mathbf{x}^*, \mathbf{x}^*)$. 

By the properties of the multivariate Gaussian distribution, the conditional distribution of $\bff^*$ given $\bff$  is also a Gaussian distribution for fixed $\mathbf{X}$ and $\mathbf{x}^*$. Extending to the general case, the conditional distribution for the latent function $\bff(\bx^*)$ is a Gaussian process
\begin{equation}
   \label{eq:conditionalGP}
\bff(\bx^*) \given \bff,\btheta  \sim \mathcal{GP}(k_{\btheta}(\bx^*,\mathbf{X})\Kff^{-1}\bff,  k_{\btheta}(\bx^*,\bx^*) -k_{\btheta}(\bx^*,\mathbf{X})\Kff^{-1}k_{\btheta}(\mathbf{X},\bx^*)).
 \end{equation}

% \begin{equation}
%    \label{eq:conditional}
% \bff^* \given \bff, \mathbf{X}, \mathbf{X}^*,\btheta  \sim \mathcal{N}(\Kfx\Kff^{-1}\bff,  \Kxx -\Kxf \Kff^{-1}\Kfx).
%  \end{equation}

 Using the modelling framework in eq. \eqref{eq:framework}, we have a Gaussian process prior $p(f \given \btheta)$ over the function $f(\mathbf{x})$. If we let $\mathcal{D} = \{\mathbf{X},\by\}$ represent our observed data, where $\mathbf{X} = (\bx_1, \bx_2, \ldots, \bx_n)$, then the likelihood of the data, conditional on function values $\bff$, is $p(\mathcal{D} \given \bff,\btheta)$. Using Bayes theorem, we can show that the posterior distribution for the function $\bff$ is $p(\bff \given \mathcal{D},\btheta) \propto p(\mathcal{D} \given \bff,\btheta)p(\bff \given \btheta)$. In the general setting, the posterior is non-Gaussian (see Section \ref{sec:noisy-gauss-proc} for an exception) and cannot be expressed in an analytic form, but can often be approximated using a Laplace approximation \citep{williamsandbarber:1998:bayesian}, expectation-propagation \citep{Minka2001}, or variational inference \citep{opper2009variational} (see \cite{Nickisch2008} for a full review). Alternatively, simulation-based inference methods including Markov chain Monte Carlo (MCMC) algorithms \citep{robert2004monte} can be applied.

From the posterior distribution, we can derive the marginal predictive distribution of $y^*$, given test points $\bx^*$, by integrating out the latent function,
\begin{equation}
  \label{eq:9}
p(y^* \given \bx^*,\mathcal{D},\btheta) = \int\int p(y^* \given f^*, \btheta)p(f^* \given \bff, \bx^*, \mathbf{X}, \btheta) p(\bff \given \mathcal{D},\btheta) df^* d\bff.  
\end{equation}
Calculating this integral is generally intractable, with the exception of nonlinear regression with Gaussian observations (see Section \ref{sec:noisy-gauss-proc}). In settings such as seen with classification models, the marginal predictive distribution is intractable, but can be approximated using the methods mentioned above. In Section \ref{sec:gauss-proc-with} we will introduce a MCMC algorithm for sampling exactly from the posterior distribution and use these samples to evaluate the marginal predictive distribution through Monte Carlo integration.

\subsection{Nonparametric regression: the analytic case}
\label{sec:noisy-gauss-proc}

We start by considering a special case of eq. \eqref{eq:framework}, where the observations follow a Gaussian distribution, 
\begin{equation}
  \label{eq:5}
y_i \sim \mathcal{N}(f(\mathbf{x}_i),\sigma^2), \ i=1,\ldots,n.  
\end{equation}
In this instance, the posterior for the latent variables, conditional on the data,  can be derived analytically as a Gaussian distribution (see \cite{rasmussen}),
 \begin{equation}
   \label{eq:6}
\bff \given \mathcal{D},\btheta  \sim \mathcal{N}(\Kff(\Kff + \sigma_n^2 \bI)^{-1}\by,  \Kff -\Kff (\Kff+ \sigma_n^2 \bI)^{-1}\Kff).
 \end{equation}
The predictive distribution for $y^*$ in eq. \eqref{eq:9} can also be calculated analytically by noting that the likelihood in eq. \eqref{eq:5}, the posterior in eq. \eqref{eq:6} and the conditional distribution for $f^*$ in eq. \eqref{eq:conditionalGP} are all Gaussian and integration over a product of Gaussians produces a Gaussian distribution, 
\begin{equation}
  \label{eq:predictive}
y^* \given \bx^*,\mathcal{D},\btheta, \sigma^2 \sim \mathcal{N}(\mu(\bx^*), \Sigma(\bx^*,\bx^{*'}) + \sigma^2\bI),  
\end{equation}
where
$$\mu(\bx^*) = k(\bx^*,\mathbf{X})(\Kff + \sigma_n^2 \bI)^{-1}\by$$
and $$\Sigma(\bx^*,\bx^{*'}) = k(\bx^*,\bx^*) -k(\bx^*,\mathbf{X})(\Kff+ \sigma_n^2 \bI)^{-1} k(\mathbf{X},\bx^*)$$
(see Chapter 2 of \cite{rasmussen} for the full derivation).

% \subsection{Maximising the marginal likelihood}
% \label{sec:maxim-marg-log}

The quality of the Gaussian process fit to the data is dependent on the model hyperparameters, $\btheta$, which are present in the mean and kernel functions as well as the observation noise $\sigma^2$. Estimating these parameters requires the marginal likelihood of the data,
\[
p(\mathcal{D} \given \btheta, \sigma) = \int p(\by \given \bff, \sigma^2)p(\bff \given \mathbf{X}, \btheta) d\bff,
\]
which is given by marginalising over the latent function values $\bff$. Assuming a Gaussian observation model in eq. \eqref{eq:5}, the marginal distribution is $p(\by \given \mathbf{X}, \btheta, \sigma^2) = \mathcal{N}(0,\Kff+\sigma^2\bI)$. For convenience of optimisation we work with the log-marginal likelihood
\begin{equation}
  \label{eq:mll}
\log p(\mathcal{D} \given \btheta, \sigma) = -\frac{1}{2}\by^\top(\Kff+ \sigma^2 \bI)^{-1}\by -\frac{1}{2}\log|\Kff + \sigma^2 \bI| -\frac{n}{2}\log 2\pi.  
\end{equation}
 The tractablility of the log-marginal likelihood allows for the straightforward calculation of the gradient with respect to the hyperparameters. Efficient gradient-based optimisation techniques (e.g. L-BFGS and conjugate gradients) can be applied to numerically maximise the log-marginal likelihood function. In practice, we utilise the excellent \pkg{Optim.jl} package \citep{Mogensen2018} and provide an interface for the user to specify their choice of optimisation algorithm. Alternatively, a Bayesian approach can be taken, where samples are drawn from the posterior of the hyperparameters using the in-built MCMC algorithm, see Section \ref{sec:package} for an example.

\subsection{Gaussian processes with non-Gaussian data}
\label{sec:gauss-proc-with}

In Section \ref{sec:noisy-gauss-proc} we considered the simple tractable case of nonlinear regression with Gaussian observations. The modelling framework given in eq. (\ref{eq:framework}) is general enough to extend the Gaussian process model to a wide range of data types. For example, Gaussian processes can be applied to binary classification problems (see \cite{rasmussen} Chapter 3), by using a Bernoulli likelihood function (see Section~\ref{sec:binary-class} for more details).

When the likelihood $p(\by \given \bff, \btheta)$ is non-Gaussian, the posterior distribution of the latent function, conditional on observed data $p(\bff \given \mathcal{D},\btheta)$, does not have a closed form solution. A popular approach for addressing this problem is to replace the posterior with an analytic approximation, such as a Gaussian distribution derived from a Laplace approximation \citep{williamsandbarber:1998:bayesian} or an expectation-propagation algorithm \citep{Minka2001}. These approximations are simple to employ and can work well in practice on specific problems \citep{Nickisch2008}, however, in general these methods struggle if the posterior is significantly non-Gaussian. Alternatively, rather than trying to find a tractable approximation to the posterior, one could sample from it and use the samples as a stochastic approximation and evaluate integrals of interest through Monte Carlo integration \citep{ripley2009stochastic}. 

Markov chain Monte Carlo methods \citep{robert2004monte} represent a general class of algorithms for sampling from high-dimensional posterior distributions. They have favourable theoretical support to guarantee algorithmic convergence \citep{Roberts2004b} and are generally easy to implement only requiring that it is possible to evaluate the posterior density pointwise. We use the centred parameterisation as given in \cite{Murray2010,Filippone2013b,Hensman2015}, which has been shown to improve the accuracy of MCMC algorithms by de-coupling the strong dependence between $\bff$ and $\btheta$. Re-parameterising eq. (\ref{eq:framework}) we have,
\begin{eqnarray}
\label{eq:framework2}
\mathbf{y} \given \bff, \btheta &\sim& \prod_{i=1}^n p(y_i\,\given \,f_i,\btheta), \nonumber \\
    \bff = L_{\btheta}\bv, &&\qquad L_{\btheta}L_{\btheta}^\top = K_{\btheta},  \\
      \bv \sim \mathcal{N}\left(\mathbf{0}_n, \bI_n\right), && \btheta \sim p(\btheta), \nonumber
\end{eqnarray}
where $L_{\btheta}$ is the lower Cholesky decomposition of the covariance matrix $K_{\btheta}$, with $(i,j)$-element $K_{i,j} = k_{\btheta}(x_i, x_j)$. The random variables $\bv$ are now independent under the prior and a deterministic transformation gives the function values $\bff$. The posterior distribution for $p(\bff \given \mathcal{D}, \btheta)$, or in the transformed setting, $p(\bv \given \mathcal{D}, \btheta)$ usually does not have a closed form expression. Using MCMC we can instead sample from this distribution, or in the case of unknown model parameters $\btheta$, we can sample from 
\begin{align}
  p(\btheta,\bv \given \mathcal{D}) \propto  p(\mathcal{D} \given \bv,\btheta)p(\bv)p(\btheta).
\end{align}

Numerous MCMC algorithms have been proposed to sample from the Gaussian process posterior (see \cite{Titsias2008} for a review). In this package we use the highly efficient Hamiltonian Monte Carlo (HMC) algorithm \citep{Neal2010}, which utilises gradient information to efficiently sample from the posterior. Under the re-parametrised model eq. \eqref{eq:framework2}, calculating the gradient of the posterior requires the derivative of the Cholesky factor $L_{\btheta}$. We calculate this derivative using the blocked algorithm of \cite{Murray2016}.

After running the MCMC algorithm we have $N$ samples $\{\bv^{(j)},\btheta^{(j)} \}_{j=1}^N$ from the posterior $p(\btheta,\bv \given \mathcal{D})$. Function values $\bff$ are given by the deterministic transform of the Monte Carlo samples, $\bff^{(i)} = L_{\btheta^{(i)}}\bv^{(i)}$. Monte Carlo integration is then used to estimate for the marginal predictive distribution from eq. (\ref{eq:9}),
\begin{equation}
  \label{eq:pred}
\hat{p}(y^* \given \bx^*,\mathcal{D},\btheta) \simeq \frac{1}{N} \sum_{i=1}^{N} \int p(y^* \given f^*, \btheta^{(i)})p(f^* \given \bff^{(i)}, \bx^*, \mathbf{X}, \btheta^{(i)}) df^*,  
\end{equation}
where we have a one-dimensional integral for $f^*$ that can be efficiently evaluated using Gauss-Hermite quadrature \citep{Liu1994}.

\subsection{Scaling Gaussian processes with sparse approximations}\label{sec:sparsegp}

When applying Gaussian processes to a dataset of size $n$, an unfortunate by-product is the $\mathcal{O}(n^3)$ scalability of the Gaussian process. This is due to the need to invert and compute the determinant of the $n\times n$ kernel matrix $\Kff$. There exist a number of approaches to deriving more scalable Gaussian processes: sparsity inducing kernels \citep{melkumyan2009sparse}, Nystr\"{o}m-based eigendecompositions \citep{williams2001using}, variational posterior approximations \citep{titsias2009variational}, neighbourhood partitioning schemes \citep{datta2016hierarchical}, and divide-and-conquer strategies \citep{guhaniyogi2017divide}. In this package, scalability within the Gaussian process model is achieved by approximating the Gaussian process' prior with a subset of inducing points $\mathbf{u}$ of size $m$, such that $m << n$ \citep{quinonero2005unifying}.

Due to the consistency of a Gaussian process\footnote{A required assumption for any valid stochastic process, consistency assumes if we marginalise out part of the process, then the resulting marginal distribution will be the same as the distribution defined in the original sequence.} the joint prior in eq. \eqref{eq:joint} can be recovered from a sparse Gaussian process through the marginalisation of $\bu$ 
\begin{equation*}
    p(\bff, \bff^*) = \int p(\bff, \bff^*, \bu) d\bu = \int p(\bff, \bff^* \given \bu)p(\bu)d\bu, 
\end{equation*}
where $\bu \sim \mathcal{N}(0,\Kuu)$ and $\Kuu= k_{\theta}(\bu, \bu)$ is an $m \times m$ covariance matrix. An approximation is only induced under the sparse framework through the assumption that $\bff$ and $\bff^*$ are conditionally independent, given $\bu$.
\begin{equation*}
    p(\bff, \bff^*) \approx q(\bff, \bff^*) = \int q(\bff \given \bu) q(\bff^* \given \bu)p(\bu)d\bu.
\end{equation*}
From this dependency structure, it can be seen that $\bff$ and $\bff^*$ are only dependent through the information expressed in $\bu$. The fundamental difference between each of the four sparse Gaussian process schemes implemented in this package is the additional assumptions that each scheme imposes upon the conditional distributions $q(\bff \given \bu)$ and $q(\bff^* \given \bu)$. In the exact case, these two conditional distributions can be expressed as 
\begin{align}
p(\bff \given \bu) & = \mathcal{N}(\Kfu \Kuu^{-1}\bu, \Kff-\Qff) \label{eq:sparse_train_cond} \\
p(\bff^* \given \bu) & = \mathcal{N}(\Kxu \Kuu^{-1}\bu, \Kxx-\Qxx)\label{eq:sparse_test_cond},
\end{align}
where $\Qff = \mathbf{K}_{\mathbf{f u}} \mathbf{K}_{\mathbf{u u}}^{-1} \mathbf{K}_{\mathbf{u f}}$. 

The simplest, and most computationally fast, sparse method is the subset of regressors (SoR)  strategy. SoR assumes a deterministic relationship between each $\bff$ and $\bu$, making the Gaussian process' marginal predictive distribution \eqref{eq:9} now equivalent to
\begin{equation}
    \label{eq:sor_conds}
    q(y^* \given \bx^*,\mathcal{D},\btheta) = \mathcal{N}\left(\sigma^{-2} \mathbf{K}_{\mathbf{f^*}, \bu} \Sigma \mathbf{K}_{\bu, \bff} \mathbf{y}, \mathbf{K}_{\mathbf{f^*}, \bu} \Sigma \mathbf{K}_{\bu, \mathbf{f^*}}\right),
\end{equation}
where $\Sigma = \left(\sigma^{-2}  \mathbf{K}_{\bu, \bff}  \mathbf{K}_{\bff, \bu}+ \mathbf{K}_{\bu, \bu}\right)^{-1}$. Such scalability comes at the great cost of wildly inaccurate predictive uncertainties that often underestimate the true posterior variance  as the model can only express $m$ degrees-of-freedom. This result occurs as at most $m$ linearly independent functions can be drawn from the prior, and consequently prior variances are poorly approximated. 

A more elegant sparse method, is the deterministic training conditional (DTC) approach of \cite{seeger2003fast}. DTC addresses the issue of inaccuracy within the Gaussian process' posterior variance by computing the Gaussian process' likelihood using information from all $n$ data points; not just $\bu$. This is achieved by projecting $\bff$ such that $\bff=\mathbf{K}_{\mathrm{f}, \mathrm{u}} \mathbf{K}_{\mathrm{u}, \mathrm{u}}^{-1} \mathrm{u}$. With an exact likelihood computation, an approximation is still required on the Gaussian process' joint prior
\begin{equation}
  \label{eq:DTCjoint}
  \left( \begin{matrix} \bff \\ \bff^* \end{matrix} \right) \given \mathbf{X}, \mathbf{x}^*,\btheta 
  \sim \mathcal{N}\left(\begin{matrix} 0, \left[\begin{matrix}
          \Qff   &  \Qfx \\
          \Qxf  &  \Kxx \\
        \end{matrix}\right]
    \end{matrix}\right).
\end{equation}
Through retention of an exact likelihood, coupled with an approximate prior, a deterministic relationship need only be imposed on $\bu$ and $\bff$, allowing for an exact test conditional (\eqref{eq:sparse_test_cond}) to be computed. Given that the test conditional is now exact, and the prior variance of $\bff^*$ is computed using $\Kxx$, not $\Qxx$, more reasonable predictive uncertainties are now produced. Note, while an exact test conditional is now being computed, a DTC approximation is not an exact Gaussian process as the process is no longer consistent across training and test cases due to the inclusion of $\Kxx$ in \eqref{eq:DTCjoint}.

The fully independent training conditional (FITC) scheme enables a richer covariance structure by preserving the exact prior covariances along the diagonal  \citep{snelson2006sparse}. This can be seen in the model's joint prior
\begin{equation}
  \label{eq:FITCjoint}
  \left( \begin{matrix} \bff \\ \bff^* \end{matrix} \right) \given \mathbf{X}, \mathbf{x}^*,\btheta 
  \sim \mathcal{N}\left(\begin{matrix} 0, \left[\begin{matrix}
          \Qff - \text{diag}\left[\Qff - \Kff \right]   &  \Qfx \\ \Qxf &  \Kxx \\
        \end{matrix}\right]
    \end{matrix}\right).
\end{equation}
As with the DTC, FITC imposes an approximation to the training conditional from \eqref{eq:sparse_train_cond}, but computes \eqref{eq:sparse_test_cond} exactly. An important extension to \eqref{eq:FITCjoint}, is proposed in \cite{quinonero2005unifying} whereby the prior variance for $\bff^*$ is reformulated as $\Qxx-\text{diag} \left[\Qxx-\Kxx \right]$. This assumption of full independence within the conditionals of both $\bff$ and $\bff^*$ ensures that the FITC approximation is equivalent to exact inference within a non-degenerate Gaussian process; a property not enjoyed by the aforementioned sparse approximations\footnote{Note, in this package \eqref{eq:FITCjoint} has been implemented, however, the proposed extension by \cite{quinonero2005unifying} is left for future work within the package}.

The final sparse method implemented within the package is the full-scale approximation of \cite{sang2012full}. A full-scale approximation further enriches the prior covariance structure by imposing a series blocked matrix corrective terms along diagonal of $\bff$
\begin{equation}
  \label{eq:FSAjoint}
  \left( \begin{matrix} \bff \\ \bff^* \end{matrix} \right) \given \mathbf{X}, \mathbf{x}^*,\btheta 
  \sim \mathcal{N}\left(\begin{matrix} 0, \left[\begin{matrix}
          \Qff - \text{blockdiag}\left[\Qff - \Kff \right]   &  \Qfx - \text{blockdiag}\left[\Qfx - \Kfx \right] \\ 
          \Qxf - \text{blockdiag}\left[\Qxf - \Kxf \right] &  \Kxx \\
        \end{matrix}\right]
    \end{matrix}\right).
\end{equation}
The predictive uncertainties that a full-scale approach yields will be far superior to any of the previous sparse approximation, however, this comes at the cost of an increased computational complexity due to a denser covariance matrix. As with the DTC and FITC approximations, the exact test conditional of eq. \eqref{eq:sparse_test_cond} is preserved, while the approximation of the training conditional in eq. \eqref{eq:sparse_train_cond} takes the form $q(\bff \given \bu) = \mathcal{N}(\Kfu \Kuu^{-1}\bu, \text{blockdiag}\left[ \Kff - \Qff  \right])$.

Adopting a full-scale approach requires the practitioner to specify the number of blocks $k$, apriori. The trade-off when making this decision is that fewer blocks will result in a more accurate predictive distribution, however, the computational complexity will increase. As recommended by \cite{tresp2000bayesian}, it is commonly advised to select $k=\frac{n}{m}$, where each block is of dimension $m \times m$. In the extreme case that $k=m$, the full-scale approach becomes a FITC approximation, and if $k=1$, just a single block will exist, and the exact Gaussian process will be recovered.

A final note with regard to sparse approximations is that the set of inducing point locations $\mathbf{X}_{\bu}$, such that $\bu = \bff(\mathbf{X}_{\bu})$, will heavily influence the process. Modern extensions to the sparse methods detailed above seek to \textit{learn} $\mathbf{X}_{\bu}$ concurrently during hyper-parameter optimisation. However, such an approach, whilst elegant, requires first-order derivatives of $\bu$ to be available; a functionality not currently available in the package. Instead, the practitioner is required to specify a set of points apriori that correspond to the coordinate values of $\mathbf{X}_{\bu}$. A simple, yet effective, approach to this is to divide the coordinate space up into equidistant knots and use these knot points as $\mathbf{X}_{\bu}$. 

% An additionally important component of the FITC approach is that, unlike SoR and DTC, the locations of each point in $\bu$ need not correspond directly to a point in $\mathcal{X}$. Instead, the location of each point in $\bu$ is \textit{learned}, such that the points' joint locations maximise the approximate Gaussian process' marginal likelihood. % The locations of the inducing points can be found jointly with the Gaussian process' hyperparameters, although this is not done within this package yet. 

\section{The package}
\label{sec:package}

The package can be downloaded from the \proglang{Julia} package repository during a \proglang{Julia} session by using the package manager tool. The \code{]} symbol activates the package manager, after which the \pkg{GaussianProcesses.jl} package can be installed with the following command \code{add GaussianProcesses}. Alternatively, the \pkg{Pkg} package can be used with command

  \begin{lstlisting}
Pkg.add("GaussianProcesses")
  \end{lstlisting}

\proglang{Julia} will also install all of the required dependency packages. Documentation for types and functions in the package, like other documentation in \proglang{Julia}, can be accessed through the help mode in the Julia REPL. Help mode is activated by typing \code{?} at the prompt, and documentation can then be searched by entering the name of a function or type.

  \begin{lstlisting}
julia> ?
help?> optimize!
search: optimize!    
  \end{lstlisting}
  \begin{lstlisting}
optimize!(gp::GPBase; kwargs...)

Optimise the hyperparameters of Gaussian process gp based on type II
maximum likelihood estimation. This function performs gradient-based
optimisation using the Optim pacakge to which the user is referred to
for further details.

Keyword arguments:

* `domean::Bool`: Mean function hyperparameters should be optimized
* `kern::Bool`: Kernel function hyperparameters should be optimized
* `noise::Bool`: Observation noise hyperparameter should be optimized (GPE only)
* `lik::Bool`: Likelihood hyperparameters should be optimized (GPMC only)
* `kwargs`: Keyword arguments for the optimize function from the Optim package
  \end{lstlisting}

The main function in the package is \code{GP}, which fits the Gaussian process model to covariates $\bX$ and responses $\by$. As discussed in the previous section, the Gaussian process is completely specified by its \code{mean} and \code{kernel} functions and possibly a \code{likelihood} when the observations $\by$ are non-Gaussian.

  \begin{lstlisting}
gp = GP(X,y,mean,kernel)
gp = GP(X,y,mean,kernel,likelihood)    
  \end{lstlisting}

This highlights the use of the \proglang{Julia} multiple dispatch feature. The \code{GP} function will, in the background, construct either an object of type \code{GPE} or \code{GPMC} for exact or Monte Carlo inference, respectively, depending on whether or not a \code{likelihood} function is specified. If no likelihood function is given, then it is assumed that $\by$ are Gaussian distributed as in the case analytic case of eq. \eqref{eq:5}. 

In this section we will highlight the functionality of the package by considering a simple Gaussian process regression example which follows the tractable case outlined in Section \ref{sec:noisy-gauss-proc}. We start by loading the package and simulating some data.

  \begin{lstlisting}
using GaussianProcesses, Random

Random.seed!(13579)             # Set the seed using the 'Random' package
n = 10;                         # number of training points
x = 2$\pi$ * rand(n);           # predictors
y = sin.(x) + 0.05*randn(n);    # regressors
  \end{lstlisting}

Note that \proglang{Julia} supports UTF-8 characters, and so one can use Greek characters to improve the readability of the code.

The first step in modelling data with a Gaussian process is to choose the mean and kernel functions which describe the data. There are a variety of mean and kernel functions available in the package (see Appendix \ref{sec:available-functions} for a list). Note that all hyperparameters for the mean and kernel functions and the observation noise, $\sigma$, are given on the log scale. The Gaussian process is represented by an object of type \code{GP} and constructed from the observation data, a mean function and kernel, and optionally the observation noise.

  \begin{lstlisting}
# Select mean and covariance function
mZero = MeanZero()                   # Zero mean function
kern = SE(0.0,0.0)                   # Sqaured exponential kernel 
logObsNoise = -1.0                   # log standard deviation of observation noise 

gp = GP(x,y,mZero,kern,logObsNoise)       # Fit the GP
  \end{lstlisting}

For this example we have used a zero mean function and squared exponential kernel with signal standard deviation and length scale parameters equal to $1.0$ (recalling that inputs are on the log scale). After fitting the \code{GP}, a summary output is produced which provides some basic information on the GP object, including the type of mean and kernel functions used, as well as returning the value of the marginal log-likelihood eq. \eqref{eq:mll}. Once the user has applied the \code{GP} function to the the data, a summary of the GP object is printed.

  \begin{lstlisting}
GP Exact object:
Dim = 1
Number of observations = 10
Mean function:
Type: GaussianProcesses.MeanZero, Params: Any[]
Kernel:
Type: GaussianProcesses.SEIso, Params: [0.0,0.0]
Input observations = 
[5.66072 1.67222 ... 6.08978 3.39451]
Output observations = [-0.505287,1.02312,0.616955,-0.777658,-0.875402,0.92976, ...
Variance of observation noise = 0.1353352832366127
Marginal Log-Likelihood = -6.719    
  \end{lstlisting}

Once we have fitted the \code{GP} function to the data we can calculate the predicted mean and variance of the function at unobserved points $\{\mathbf{x}^\ast,y^\ast\}$, conditional on the observed data $\mathcal{D}=\{\mathbf{y},\mathbf{X}\}$. This is done with the \code{predict\_y} function. We can also calculate the predictive distribution for the latent function $f^*$ using the \code{predict\_f} function. The \code{predict\_y} function returns the mean vector $\mu(\mathbf{x}^\ast)$ and covariance matrix $\Sigma(\mathbf{x}^\ast,\mathbf{x}^{\ast^\prime})$  of the predictive distribution in eq. \eqref{eq:predictive} (or variance vector if \code{full\_cov=false}).

  \begin{lstlisting}
x = 0:0.1:2$\pi$             # a sequence between 0 and 2$\pi$ with 0.1 spacing
$\mu$, $\Sigma$ = predict_y(gp,x);  
  \end{lstlisting}

Plotting one and two-dimensional GPs is straightforward and in the package we utilise the recipes approach to plotting graphs from the \pkg{Plots.jl}\footnote{http://docs.juliaplots.org/latest/} package. \pkg{Plots.jl} provides a general interface for plotting with several different backends including \pkg{PyPlot.jl}\footnote{https://github.com/JuliaPy/PyPlot.jl}, \pkg{Plotly.jl}\footnote{https://plot.ly/julia/} and \pkg{GR.jl}\footnote{https://github.com/jheinen/GR.jl}. The default plot function \code{plot(gp)} outputs the predicted mean and variance of the function (i.e. uses \code{predict\_f} in the background), with the uncertainty in the function represented by a confidence ribbon (set to 95\% by default). All optional plotting arguments are given after the \code{;} symbol.

  \begin{lstlisting}
using Plots
pyplot()   # Optionally select a plotting backend
# Plot the GP
plot(gp; xlabel="x", ylabel="y", title="Gaussian process", legend=false, fmt=:png)  
  \end{lstlisting}

\begin{figure}[!h]
  \centering
  \includegraphics[width=.49\textwidth]{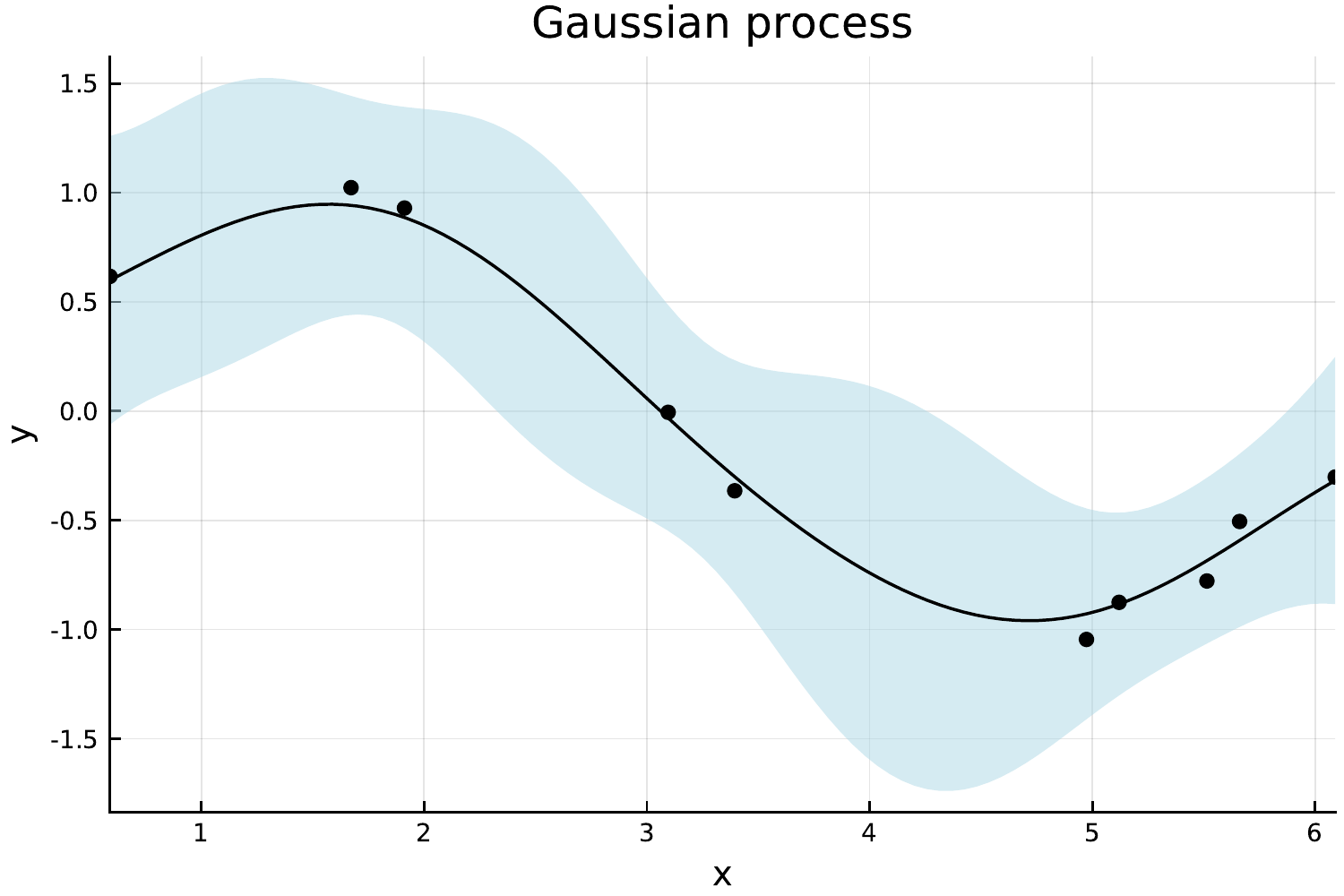}
  \includegraphics[width=.49\textwidth]{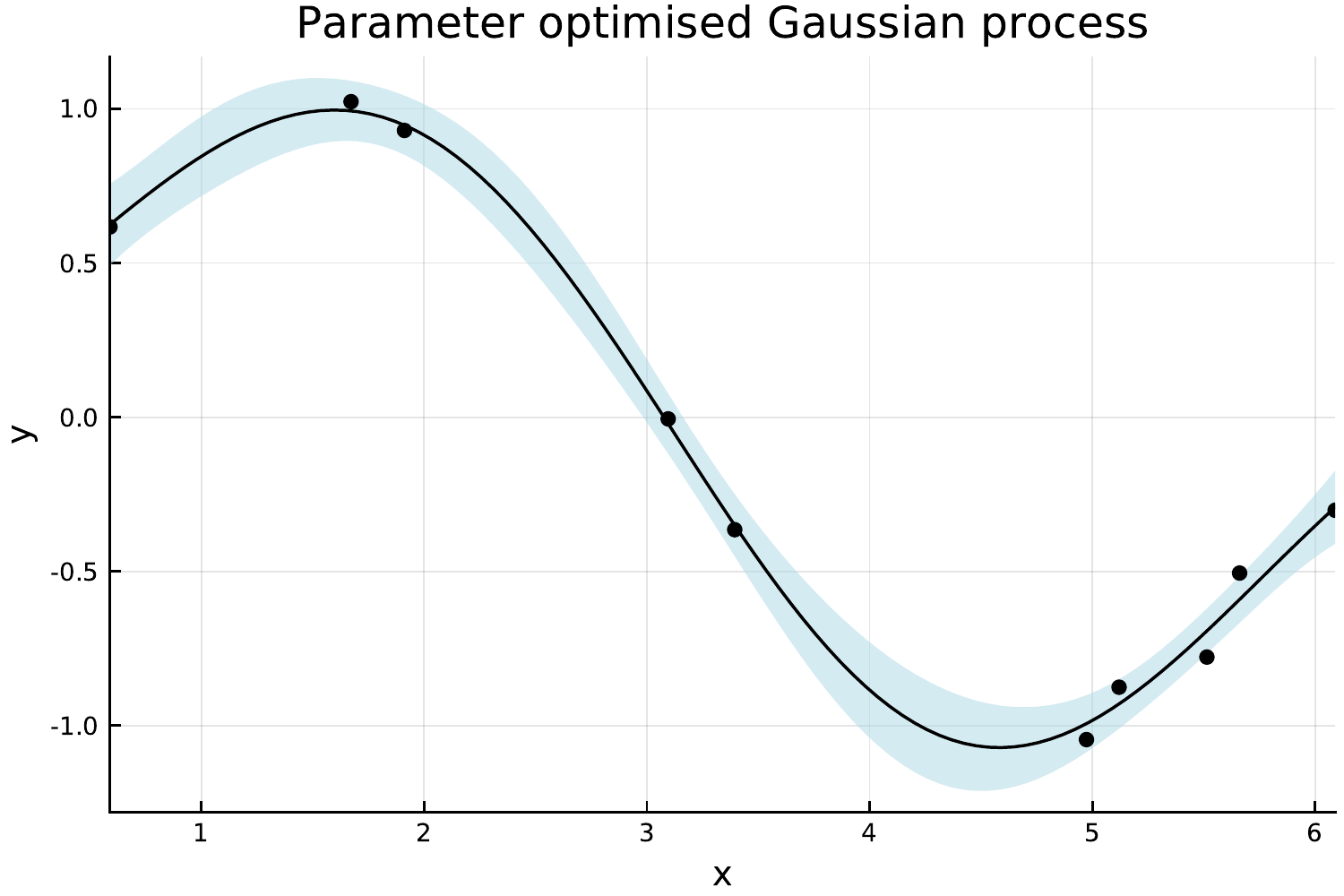}
  \label{fig:gp}
  \caption{One dimensional Gaussian process regression with initial kernel parameters (left) and optimised parameters (right).}
\end{figure}

The parameters $\btheta$ are optimised using the \pkg{Optim.jl} package (see the right-hand side of Figure \ref{fig:gp}). This offers users a range of optimisation algorithms which can be applied to estimate the parameters using maximum likelihood estimation. Gradients are available for all mean and kernel functions used in the package and therefore it is recommended that the user utilises gradient-based optimisation techniques. As a default, the \code{optimize!} function uses the L-BFGS solver, however, alternative solvers can be applied and the user should refer to the \pkg{Optim.jl} documentation for further details.

  \begin{lstlisting}
optimize!(gp)   #Optimise the parameters    
  \end{lstlisting}
  \begin{lstlisting}
Results of Optimization Algorithm
* Algorithm: L-BFGS
* Starting Point: [-1.0,0.0,0.0]
* Minimizer: [-2.683055260944582,0.4342151847965596, ...]
* Minimum: -4.902989e-01
* Iterations: 9
* Convergence: true
* |x - x'| < 1.0e-32: false
* |f(x) - f(x')| / |f(x)| < 1.0e-32: false
* |g(x)| < 1.0e-08: true
* f(x) > f(x'): false
* Reached Maximum Number of Iterations: false
* Objective Function Calls: 38
* Gradient Calls: 38
  \end{lstlisting}

Parameters can be estimated using a Bayesian approach, where instead of maximising the log-likelihood function, we can approximate the marginal posterior distribution $p(\btheta,\sigma \given \mathcal{D}) \propto p(\mathcal{D} \given \btheta,\sigma)p(\btheta,\sigma)$. We use MCMC sampling (specifically HMC sampling) to draw samples from the posterior distribution with the \code{mcmc} function. Prior distributions are assigned to the parameters of the mean and kernel parameters through the \code{set\_priors!} function. The log-noise parameter $\sigma$ is set to a non-informative prior $p(\sigma) \propto 1$. A wide range of prior distributions are available through the \pkg{Distributions.jl} package. Further details on the MCMC sampling of the package is given in Section \ref{sec:binary-class}.

  \begin{lstlisting}
using Distributions
# Uniform priors are used as default if priors are not specified
set_priors!(kern, [Normal(0,1), Normal(0,1)]) 
chain = mcmc(gp)
plot(chain', label=["Noise", "SE log length", "SE log scale"])    
  \end{lstlisting}

\begin{figure}[h]
  \centering
  \label{fig:mcmc_out}
  \includegraphics[scale=0.4]{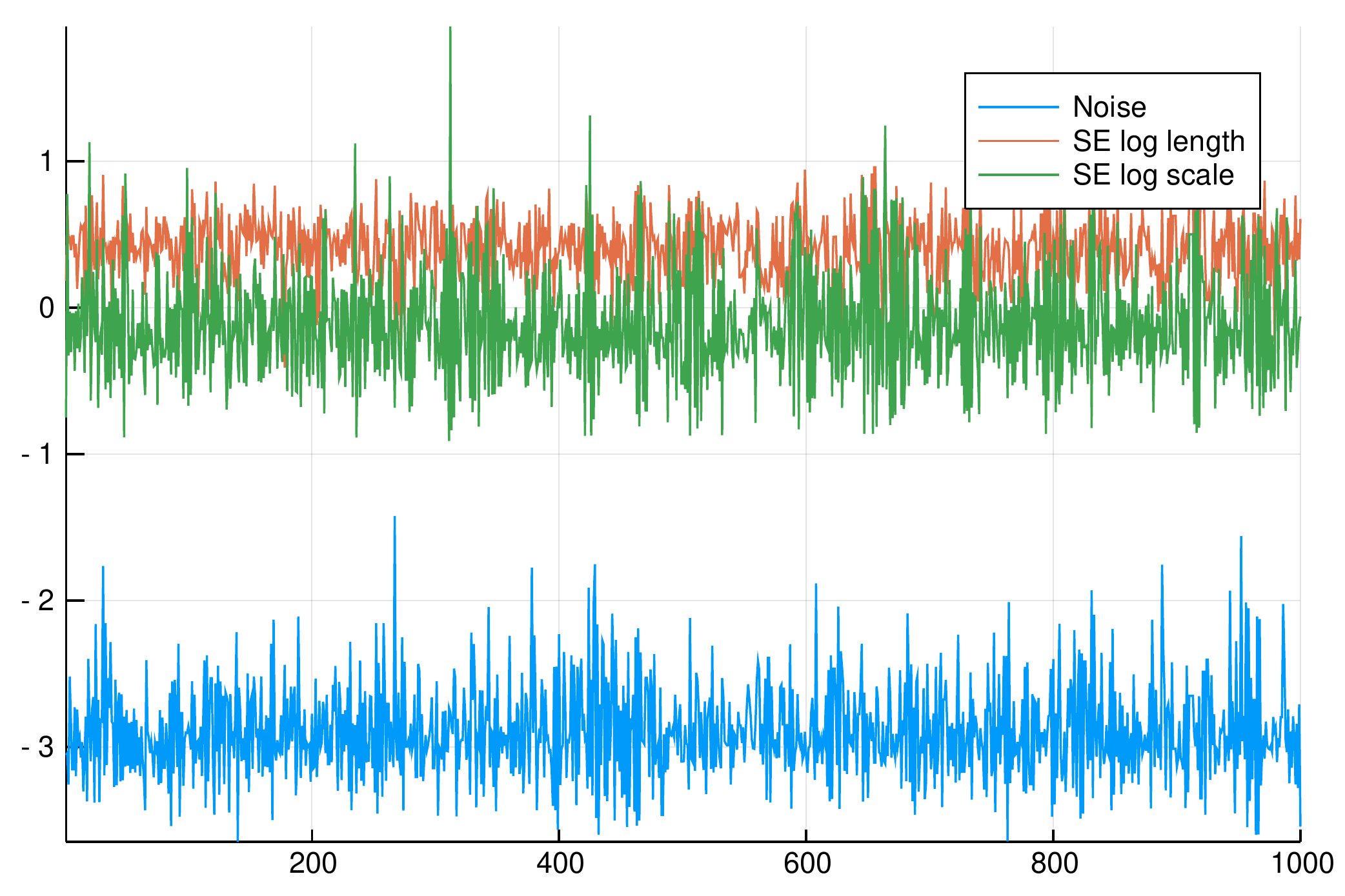}
  \caption{Trace plots of the MCMC output for the posterior samples}
\end{figure}

% \todo{DECIDE WHAT TO DO WITH THIS PART ON SAMPLING FROM GPS}
% \begin{lstlisting}
% plot(gp)                  #Plot the GP after the hyperparameters have been optimised 
% samples = rand(gp, x, 5)  #Sample from the GP
% plot!(x, samples)         #Overlay samples using the bang function
% \end{lstlisting}

% \begin{figure}[h]
%   \centering
%   \includegraphics[scale=0.5]{1d-regression-optimized.png}
% \end{figure}

The regression example above can be easily extended to higher dimensions. For the purpose of visualisation, and without loss of generality, we consider a two-dimensional regression example. When $d>1$ (recalling that $\mathcal{X} \subseteq \mathbb{R}^d$), there is the option to either use an isotropic (Iso) kernel or an automatic relevance determination (ARD) kernel. The Iso kernels have one length scale parameter $\ell$ which is the same for all dimensions. The ARD kernels, however, have different length scale parameters for each dimension. To obtain Iso or ARD kernels, a kernel function is called either with a single length scale parameter or with a vector of parameters. For example, below we will use the Mat\'ern 5/2 ARD kernel, if we wanted to use the Iso alternative instead, we would set the kernel as \code{kern=Matern(5/2,0.0,0.0)}.

In this example we use a composite kernel represented as the sum of a Mat\'ern 5/2 ARD kernel and a squared exponential isotropic kernel. This is easily implemented using the \code{+} symbol, or in the case of a product kernel, using the \code{*} symbol.

  \begin{lstlisting}
# Simulate data for a 2D Gaussian process
n = 10                 # number of data points
X = 2π * rand(2, n)    # inputs  
y = sin.(X[1,:]) .* cos.(X[2,:]) + 0.5 * rand(n) # outputs

kern = Matern(5/2,[0.0,0.0],0.0) + SE(0.0,0.0)  # sum of two kernels
   
gp2 = GP(X,y,MeanZero(),kern)
    
 \end{lstlisting}
  \begin{lstlisting}
GP Exact object:
Dim = 2
Number of observations = 10
Mean function:
Type: GaussianProcesses.MeanZero, Params: Any[]
Kernel:
 Type: GaussianProcesses.SumKernel
 Type: GaussianProcesses.Mat52Ard, Params: [-0.0, -0.0, 0.0]
 Type: GaussianProcesses.SEIso, Params: [0.0, 0.0]
Input observations = 
[5.28142 6.07037 ... 2.27508 0.15818; 3.72396 2.72093 ... 3.54584 4.91657]
Output observations = [1.03981, 0.427747, -0.0330328, 1.0351, 0.889072, 0.491157, ...
Variance of observation noise = 0.01831563888873418
Marginal Log-Likelihood = -12.457    
 \end{lstlisting}

\begin{figure}[h]
  \centering
  \label{fig:meanVar}
  \includegraphics[scale=0.5]{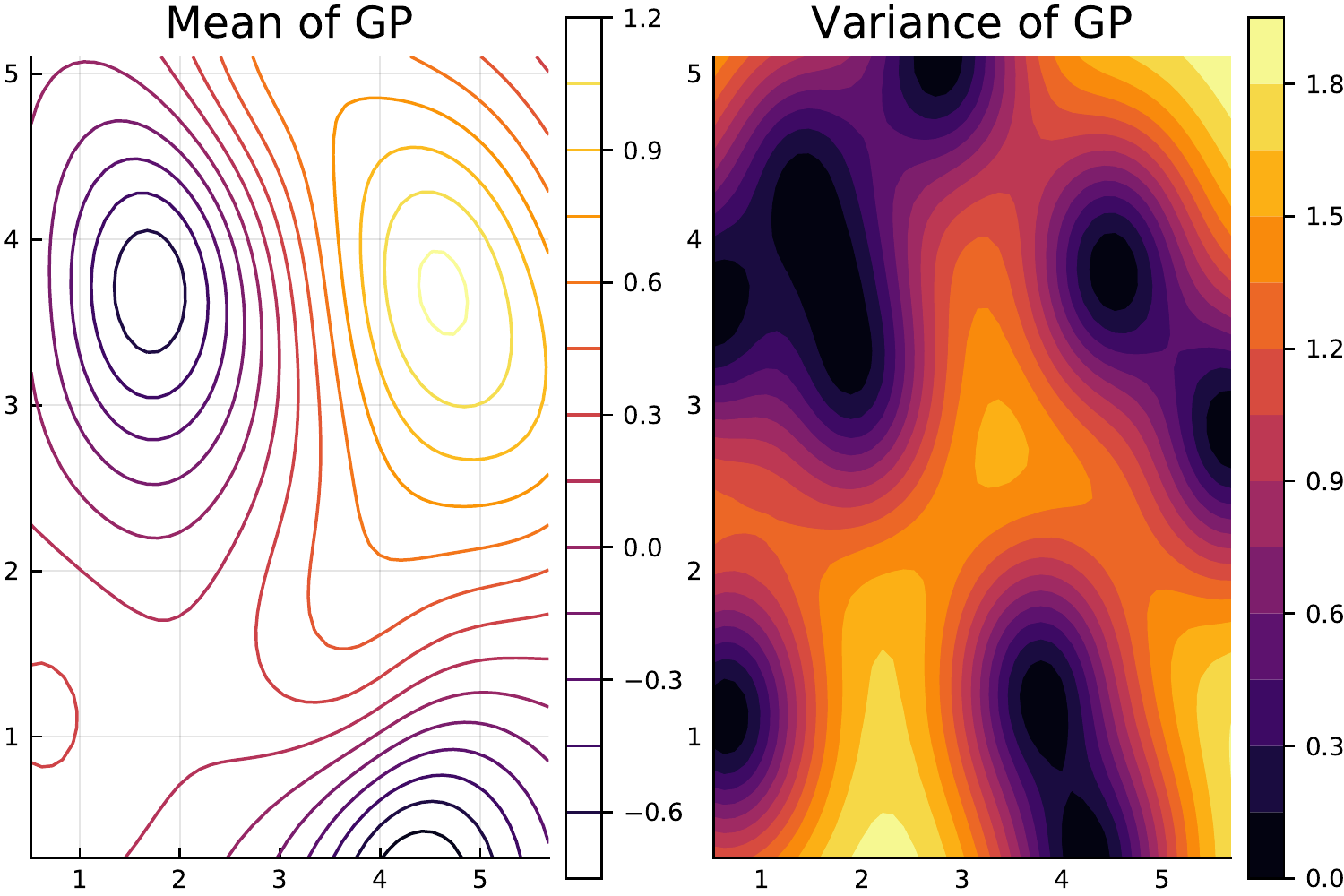}
  \caption{GP mean and variance from the two-dimensional process.}
\end{figure}

By default, the in-built \code{plot} function returns only the mean of the GP in the two-dimensional setting. There is an optional \code{var} argument which can be used to plot the two-dimensional variance (see Figure \ref{fig:meanVar}).

  \begin{lstlisting}
# Plot mean and variance
p1 = plot(gp2; title="Mean of GP")
p2 = plot(gp2; var=true, title="Variance of GP", fill=true)
plot(p1, p2; fmt=:pdf)    
  \end{lstlisting}

The \pkg{Plots.jl} package provides a flexible recipe structure which allows the user to change the plotting backend, e.g. \pkg{PyPlot.jl} to \pkg{GR.jl}. The  package also provides a rich array of plotting functions, such as contour, surface and heatmap plots.

  \begin{lstlisting}
gr() # use GR backend to allow wireframe plot
p1 = contour(gp2)
p2 = surface(gp2)
p3 = heatmap(gp2)
p4 = wireframe(gp2)
plot(p1, p2, p3, p4; fmt=:pdf)    
  \end{lstlisting}

\begin{figure}[h]
  \centering
  \label{fig:variedPlots}
  \includegraphics[scale=0.5]{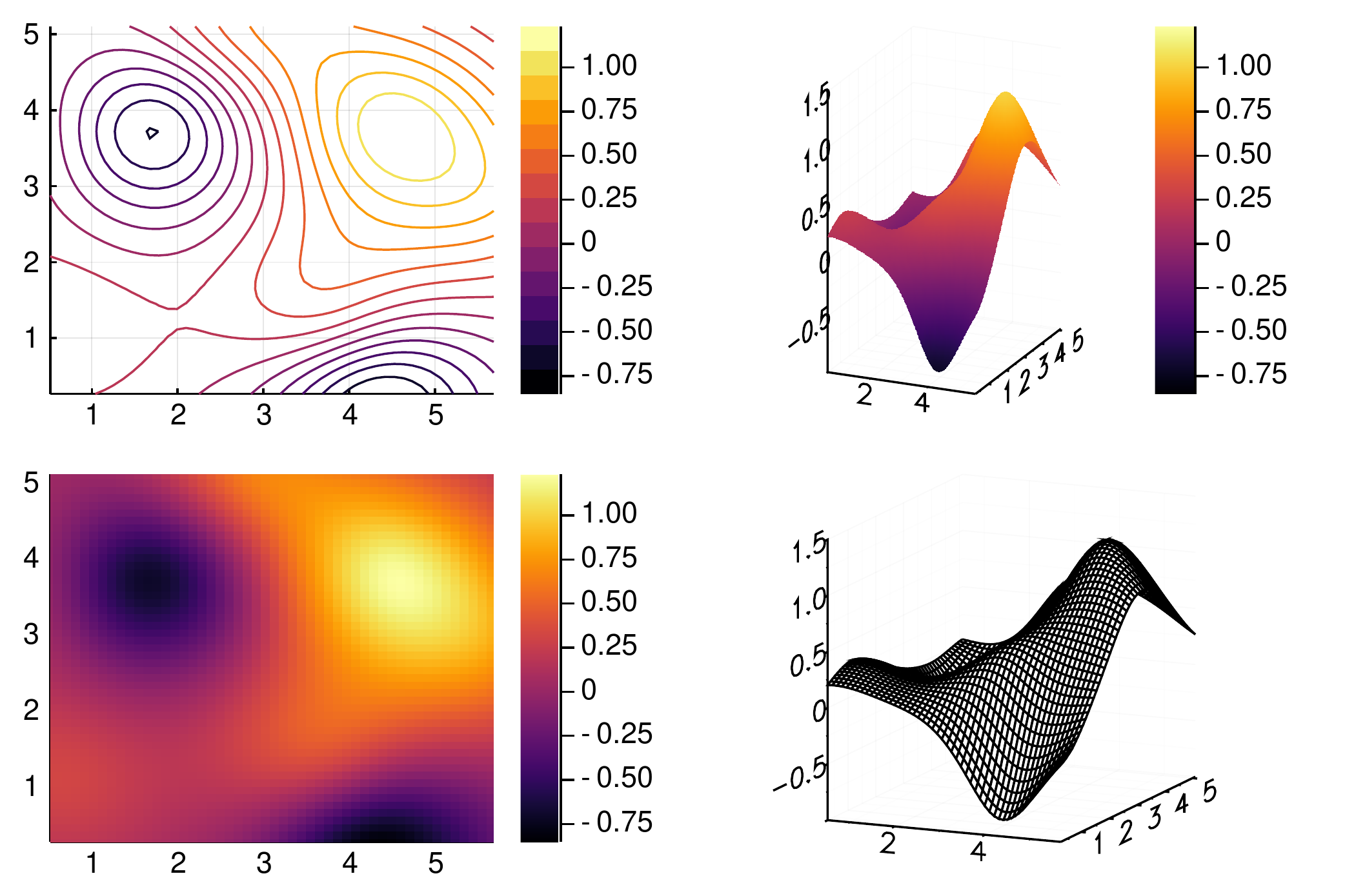}
  \caption{Two-dimensional plot of the GP mean with a range of available plotting options. Clockwise from the top left: contour, surface, wireframe and heatmap plots}
\end{figure}

\section{Demos}
\label{sec:illustration}

So far we have considered Gaussian processes where the data $\by$ are assumed to follow a Gaussian distribution centred around the latent Gaussian process function $\bff$ eq. \eqref{eq:5}. As highlighted in Section \ref{sec:gauss-proc-with}, Gaussian processes can easily be extended to model non-Gaussian data by assuming that the data are conditional on a latent Gaussian process function. This approach has been widely applied, for example, in machine learning for classification problems \citep{williams1998bayesian} and in geostatistics for spatial point process modelling \citep{moller1998log}. In this section, we will show how the \pkg{GaussianProcesses.jl} package can be used to fit Gaussian process models for binary classification, time series and count data.

\subsection{Binary classification}
\label{sec:binary-class}

In this example we show how the GP Monte Carlo function can be used for supervised learning classification. We use the Crab dataset from the \proglang{R} package \pkg{MASS}. In this dataset we are interested in predicting whether a crab is of colour form blue or orange. Our aim is to perform a Bayesian analysis and calculate the posterior distribution of the latent GP function $\bff$ and parameters $\btheta$ from the training data $\{\bX,\by\}$.

  \begin{lstlisting}
using GaussianProcesses, RDatasets, Random

Random.seed!(113355)                     

crabs = dataset("MASS","crabs");              # load the data 
crabs = crabs[shuffle(1:size(crabs)[1]), :];  # shuffle the data

train = crabs[1:div(end,2),:];             # training data

y = Array{Bool}(undefsize(train)[1]);      # response
y[train[:Sp].=="B"]=0;                     # convert characters to booleans
y[train[:Sp].=="O"]=1;

X = convert(Array,train[:,4:end]);         # predictors
  \end{lstlisting}

We assume a zero mean GP with a Mat\'ern 3/2 kernel. We use the automatic relevance determination (ARD) kernel to allow each dimension of the predictor variables to have a different length scale. As this is binary classification, we use the Bernoulli likelihood,
\[
y_i \sim \mbox{Bernoulli}(\Phi(f_i)),
\]
where $\Phi: \mathbb{R} \rightarrow [0,1]$ is the cumulative distribution function of a standard Gaussian and acts as a link function that maps the GP function to the interval [0,1], giving the probability that $y_i=1$.
Note that \code{BernLik} requires the observations to be of type \code{Bool} and unlike some likelihood functions (e.g. Student-t) does not contain any parameters to be set at initialisation.

  \begin{lstlisting}
#Select mean, kernel and likelihood function
mZero = MeanZero();                  # Zero mean function
kern = Matern(3/2,zeros(5),0.0);     # Matern 3/2 ARD kernel 
lik = BernLik();                     # Bernoulli likelihood for binary data {0,1}    
  \end{lstlisting}

We fit the Gaussian process using the general \code{GP} function. This function is a shorthand for the \code{GPMC} function, which is used to generate Monte Carlo approximations of the latent function when the likelihood is non-Gaussian.

  \begin{lstlisting}
gp = GP(X',y,mZero,kern,lik)      # Fit the Gaussian process model  
  \end{lstlisting}
  \begin{lstlisting}
GP Monte Carlo object:
Dim = 5
Number of observations = 100
Mean function:
Type: GaussianProcesses.MeanZero, Params: Float64[]
Kernel:
Type: GaussianProcesses.Mat32Ard, Params: [-0.0,-0.0,-0.0,-0.0,-0.0,0.0]
Likelihood:
Type: GaussianProcesses.BernLik, Params: Any[]
Input observations = 
[16.2 11.2 ... 11.6 18.5; 13.3 10.0 ... 9.1 14.6; ... ; 41.7 26.9 ... 28.4 42.0;
Output observations = Bool[false,false,false,false,true,true,false,true, ...  
Log-posterior = -161.209  
  \end{lstlisting}

As we are taking a Bayesian approach to infer the latent function and model parameters, we shall assign prior distributions to the unknown variables. As outlined in the general modelling framework \eqref{eq:framework2}, the latent function $\bff$ is reparameterised as $\bff=L_{\btheta}\bv$, where $\bv \sim \mathcal{N}\left(\mathbf{0}_n, \bI_n\right)$ is the prior on the transformed latent function. Using the \pkg{Distributions.jl} package we can assign normal priors to each of the Mat\'ern kernel parameters. If the mean and likelihood functions also contained parameters then we could set these priors in the way same using \code{gp.mean} and \code{gp.lik} in place of \code{gp.kernel}, respectively.

  \begin{lstlisting}
set_priors!(gp.kernel,[Distributions.Normal(0.0,2.0) for i in 1:6])    
  \end{lstlisting}

Samples from the posterior distribution of the latent function and parameters $\bff, \theta \given \mathcal{D}$, are drawn using MCMC sampling. The \code{mcmc} function uses a Hamiltonian Monte Carlo sampler \citep{Neal2010}. By default, the function runs for \code{nIter=1000} iterations and uses a step-size of $\epsilon=0.01$ with a random number of leap-frog steps $L$ between 5 and 15. Setting \code{Lmin=1} and \code{Lmax=1} gives the MALA algorithm \citep{Roberts1998}. Additionally, after the MCMC sampling is complete, the Markov chain can be post-processed using the \code{burn} and \code{thin} arguments to remove the burn-in phase (e.g. first 100 iterations) and thin the Markov chain to reduce the autocorrelation by removing values systematically (e.g. if \code{thin=5} then only every fifth value is retained).

  \begin{lstlisting}
samples = mcmc(gp; $\epsilon$=0.01, nIter=10000, burn=1000, thin=10);  
  \end{lstlisting}

We assess the predictive accuracy of the fitted model against a held-out test dataset

  \begin{lstlisting}
test = crabs[div(end,2)+1:end,:];          # select test data
yTest = Array{Bool}(undef,size(test)[1]);  # test response data
yTest[test[:Sp].=="B"]=0;                  # convert characters to booleans
yTest[test[:Sp].=="O"]=1;
xTest = convert(Array,test[:,4:end]);    
  \end{lstlisting}

Using the posterior samples $\{\bff^{(i)},\btheta^{(i)}\}^N_{i=1}$ from $p(\bff,\btheta \given \mathcal{D})$ we can make predictions about $y^*$, as in eq. \eqref{eq:pred}, using the \code{predict\_y} function and sample predictions conditional on the MCMC samples. We do this by looping over the $N$ posterior samples and for each iteration $i$ we fix the GP function $\bff^{(i)}$ and hyperparameters $\btheta^{(i)}$ to their posterior sample value.

  \begin{lstlisting}
ymean = Array{Float64}(undef,size(samples,2),size(xTest,1));

for i in 1:size(samples,2)
   set_params!(gp,samples[:,i])          # Set the GP parameters
                                            to the posterior values
   update_target!(gp)                    # Update the GP function
                                            with the new parameters
   ymean[i,:] = predict_y(gp,xTest')[1]  # Store the predictive mean
end    
 \end{lstlisting}

\begin{figure}[h]
  \centering
  \includegraphics[width=.6\textwidth]{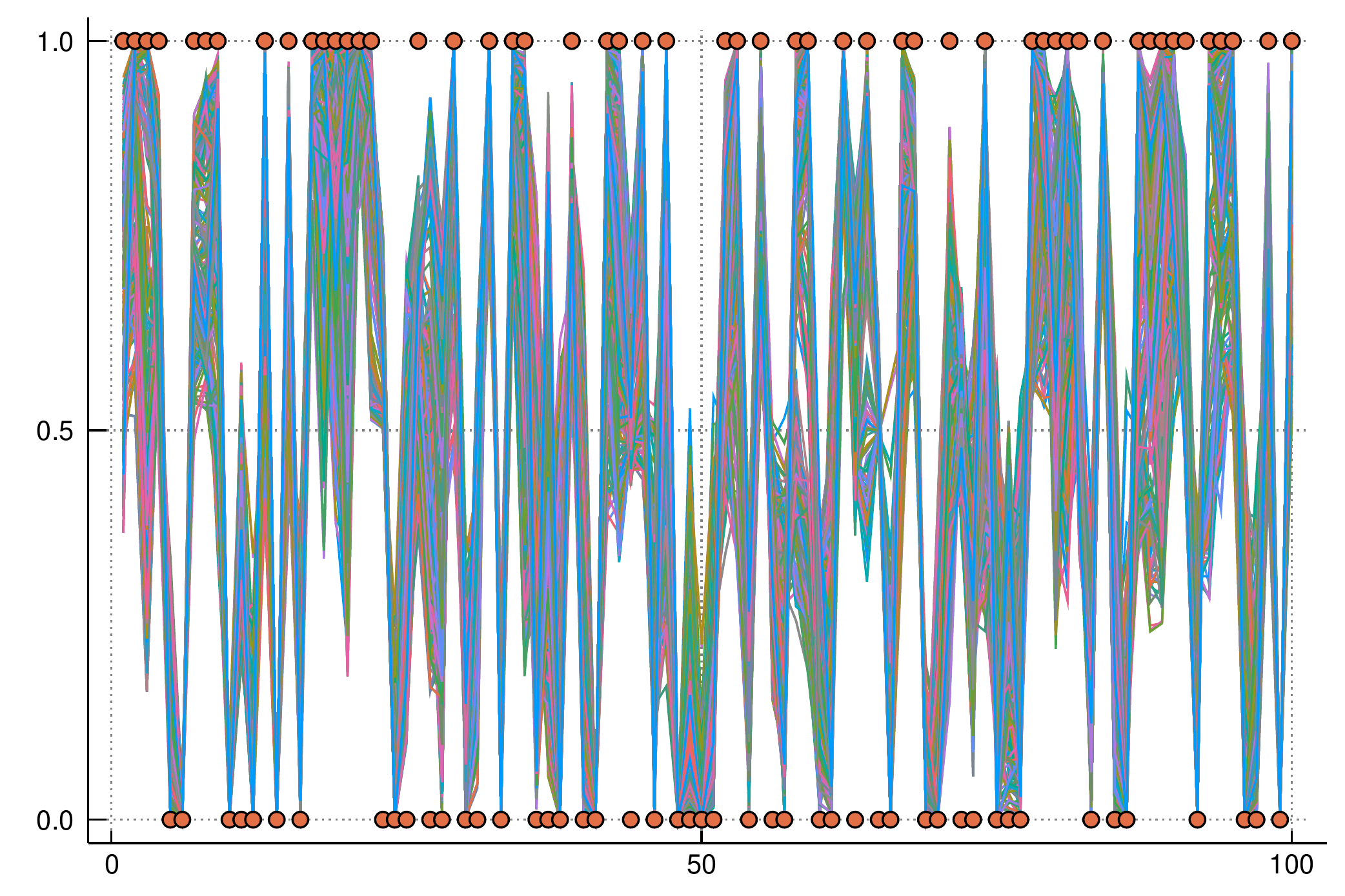}
  \caption{}
  \label{fig:crabs}
\end{figure}

For each of the posterior samples we plot (see Figure \ref{fig:crabs}) the predicted observation $y^*$  (given as lines) and overlay the true observations from the held-out data (circles).

  \begin{lstlisting}
using Plots
gr()

plot(ymean',leg=false)
scatter!(yTest)    
  \end{lstlisting}

\subsection{Time series}
\label{sec:time-series}

Gaussian processes can be used to model nonlinear time series. We consider the problem of predicting future concentrations of CO$_2$ in the atmosphere. The data are taken from the Mauna Loa observatory in Hawaii which records the monthly average atmospheric concentration of CO$_2$ (in parts per million) between 1958 to 2015. For the purpose of testing the predictive accuracy of the Gaussian process model, we fit the GP to the historical data from 1958 to 2004 and optimise the parameters using maximum likelihood estimation. 

We employ a seemingly complex kernel function to model these data which follows the kernel structure given in \cite[Chapter 5]{rasmussen}. The kernel comprises of simpler kernels with each kernel term accounting for a different aspect in the variation of the data. For example, the \code{Periodic} kernel captures the seasonal effect of CO$_2$ absorption from plants. A detailed description of each kernel contribution is given in \cite[Chapter 5]{rasmussen}.

  \begin{lstlisting}
using GaussianProcesses, DelimitedFiles

data = readdlm("data/CO2_data.csv",',')

year = data[:,1]; co2 = data[:,2];
# Split the data into training and testing data
xtrain = year[year.<2004]; ytrain = co2[year.<2004];
xtest = year[year.>=2004]; ytest = co2[year.>=2004];

# Kernel is represented as a sum of kernels
kernel = SE(4.0,4.0) + Periodic(0.0,1.0,0.0) * SE(4.0,0.0)
            + RQ(0.0,0.0,-1.0) + SE(-2.0,-2.0);

gp = GP(xtrain,ytrain,MeanZero(),kernel,-2.0)   #Fit the GP

optimize!(gp) #Estimate the parameters through maximum likelihood estimation

$\mu$, $\Sigma$ = predict_y(gp,xtest);

using Plots    #Load the Plots.jl package with the pyplot backend
pyplot()

plot(xtest,$\mu$,ribbon=$\Sigma$, title="Time series prediction",
     label="95% predictive confidence region",fmt=:pdf)
scatter!(xtest,ytest,label="Observations")    
  \end{lstlisting}

\begin{figure}[h]
  \centering
  \includegraphics[scale=0.7]{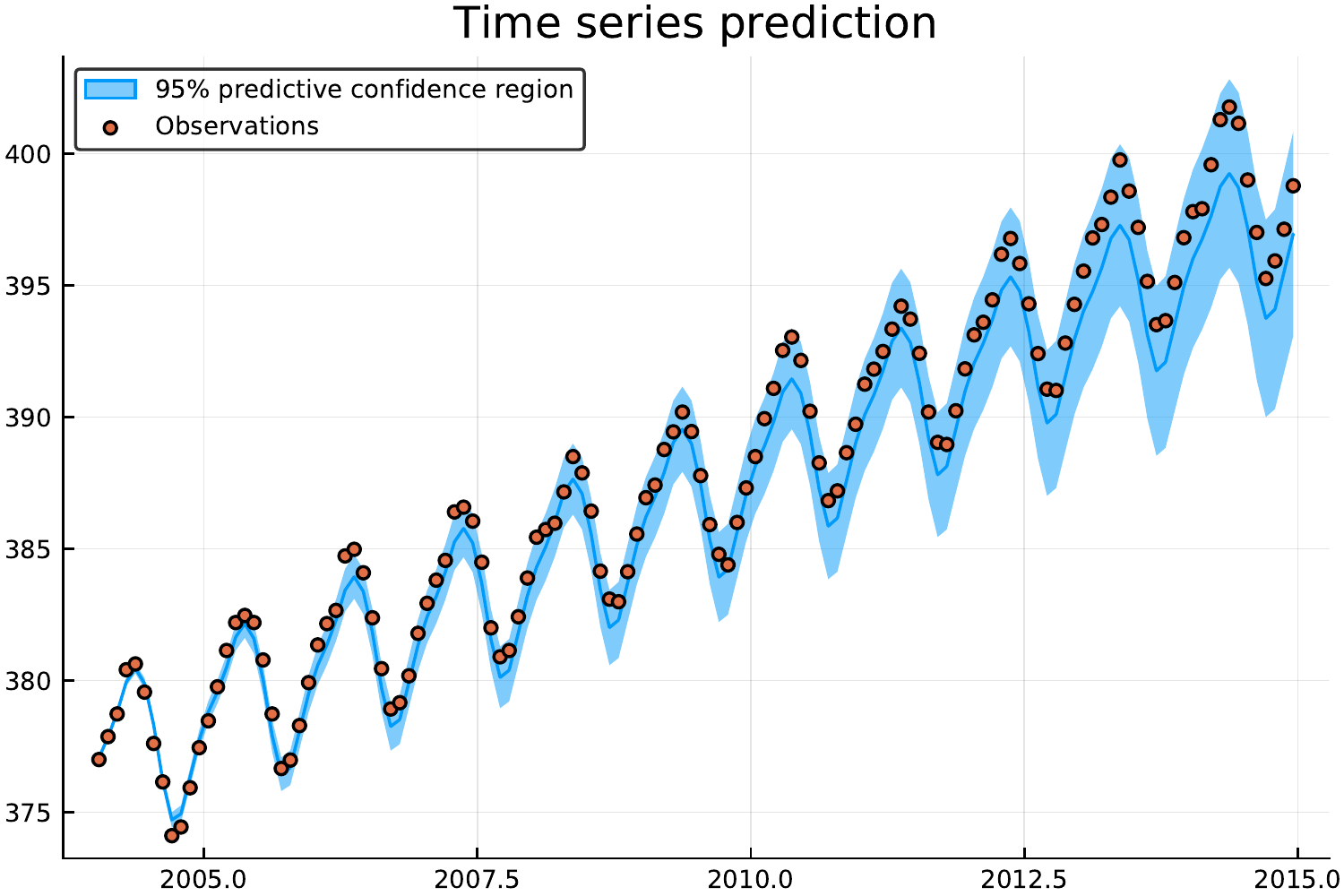}
  \caption{Predictive mean and 95\% confidence interval for CO$_2$ measurements at the Mauna Loa observatory from 2004 to 2015}
  \label{fig:time_series}
\end{figure}

The predictive accuracy of the Gaussian process is plotted in Figure \ref{fig:time_series}. Over the ten year prediction horizon the GP is able to accurately capture both the trend and seasonal variations of the CO$_2$ concentrations. Arguably, the GP prediction gradually begins to underestimate the CO$_2$ concentration. The accuracy of the fit could be further improved by extending the kernel function to include additionally terms. Recent work on automatic structure discovery \citep{Duvenaud2013a} could be used to optimise the modelling process.

\subsection{Count data}
\label{sec:count-data}

Gaussian process models can be incredibly flexible for modelling non-Gaussian data. One such example is in the case of count data $\mathbf{y}$, which can be modelled with a Poisson distribution, where the log-rate parameter can be modelled with a latent Gaussian process.
$$
\mathbf{y}  \given  \bff \sim \prod_{i=1}^{n} \frac{\lambda_i^{y_i}\exp\{-\lambda_i\}}{y_i!},
$$
where $\lambda_i=\exp(f_i)$ and $f_i$ is the latent Gaussian process.

In this example we will consider the dataset of recorded annual British coal mining disasters between 1851 and 1962. These data have been analysed previously \citep{Adams2009,Lloyd2014} and it has been shown that they follow a non-homogeneous Poisson process. These data have also been extensively analysed in the changepoint literature \cite{Carlin1992,Fearnhead2006} to identify the year in which there is a structural change to the data, i.e. change in Poisson intensity. We fit a Gaussian process to the coal mining data using a Poisson likelihood function with a Mat\'ern $3/2$ kernel function. MCMC is used to sample from the posterior distribution of the latent function and kernel parameters.

  \begin{lstlisting}
using GaussianProcesses, Distributions, Plots
pyplot()

coal = readcsv("notebooks/data/coal.csv")

X = coal[:,1]; Y = convert(Array{Int},coal[:,2]); 

#GP set-up
k = Matern(3/2,0.0,0.0)         # Matern 3/2 kernel
l = PoisLik()                   # Poisson likelihood
gp = GP(X, Y, MeanZero(), k, l) # Fit the GP

#Set the priors and sample from the posterior
set_priors!(gp.kernel,[Normal(-2.0,4.0),Normal(-2.0,4.0)]) 
samples = mcmc(gp; ε=0.08, nIter=11000,burn=1000,thin=10);

#Sample posterior function realisations
x = linspace(minimum(gp.X),maximum(gp.X),50);
fsamples = Array{Float64}(undef,size(samples,2), length(x));
for i in 1:size(samples,2)
   set_params!(gp,samples[:,i])
   update_target!(gp)
   fsamples[i,:] = rand(gp, x)
end  
  \end{lstlisting}

\begin{figure}[h]
  \centering
  \includegraphics[width=.49\textwidth]{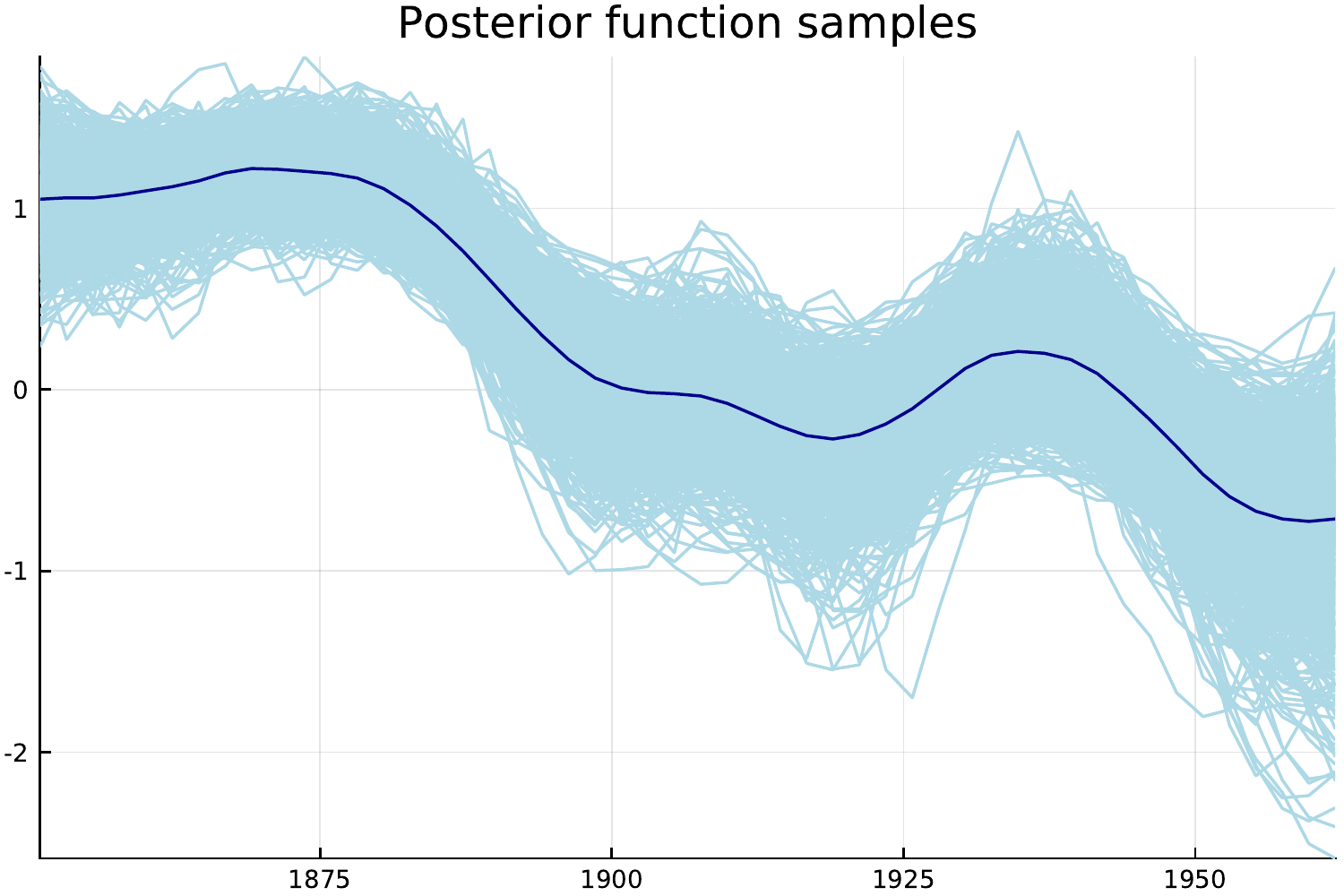}
  \includegraphics[width=.49\textwidth]{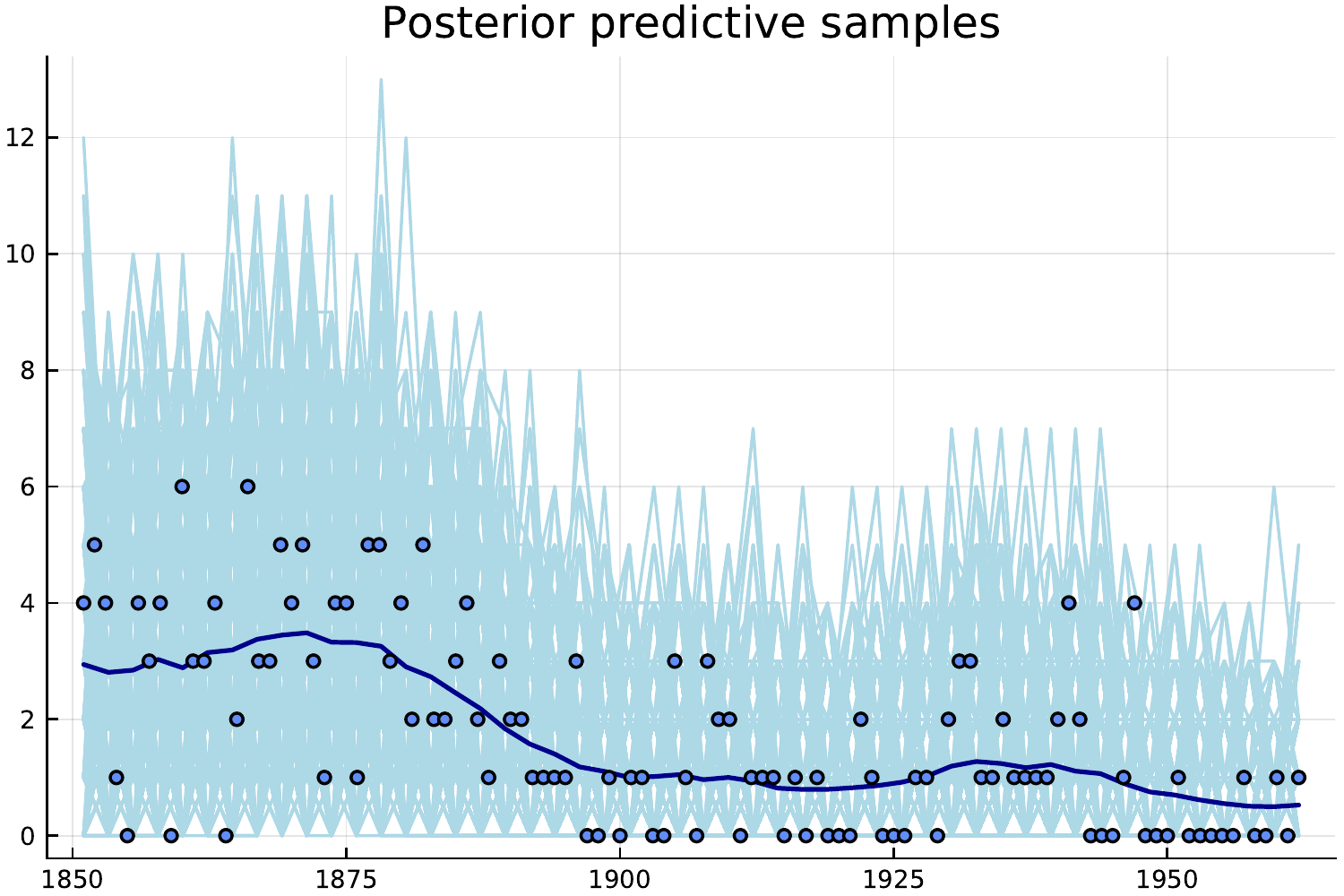}
  \caption{Samples from the posterior (left) and predictive distributions (right) of the Gaussian process with a Poisson observation model. The points show the recorded number of incidents and the dark blue lines represent the mean posterior and predictive functions, respectively.}
  \label{fig:poisson}
\end{figure}

A visual inspection, given in Figure \ref{fig:poisson}, of the latent function reveals a change in the log-intensity of the Poisson process after the year 1876. This change corresponds with several pieces of parliamentary legislation in the UK between 1870-1890 intended to improve the safety standards in British coal mines. Additionally, there appears to be a further decline in the number of accidents after 1935.

\subsection{Bayesian optimization}
\label{sec:bayes-optim}

This section introduces the \pkg{BayesianOptimization.jl}\footnote{https://github.com/jbrea/BayesianOptimization.jl} package, which requires \pkg{GaussianProcesses.jl} as a dependency. We highlight some of the memory-efficiency features of \proglang{Julia} and show how Gaussian processes can be applied to optimize noisy or costly black-box objective functions \cite{Shahriari16}.
In Bayesian optimization, an objective function $l(\bx)$ is evaluated at some points $y_1 =l(\bx_1), y_2 = l(\bx_2), \ldots, y_t = l(\bx_t)$.
A model $\mathcal M(\mathcal D_t)$, e.g. a Gaussian process, is fitted to these observations $\mathcal D_t = \{(\bx_i, y_i)\}_{i=1,\ldots,t}$ and used to determine the next input point $\bx_{t+1}$ at which the objective function should be evaluated.
The model is refitted with inclusion of the new observation $(\bx_{t+1}, y_{t+1})$ and $\mathcal M(\mathcal D_{t+1})$ is used to acquire the next input point.
With a clever acquisition of next input points, Bayesian optimization can find the optima of the objective function with fewer function evaluations than alternative optimization methods \cite{Shahriari16}.

Since the observed data sets in different time steps are highly correlated, $\mathcal D_{t+1} = \mathcal D_t \cup \{(\bx_{t+1}, y_{t+1})\}$, it would be wasteful to refit a Gaussian process to $\mathcal D_{t+1}$ without considering the model $\mathcal M(\mathcal D_t)$ that was already fit to $\mathcal D_t$.
To avoid refitting, \pkg{GaussianProcesses.jl} includes the function \code{ElasticGPE} that creates a Gaussian process where it costs little to add new observations.
In the following example, we create an elastic and exact GP for two input dimensions with an initial capacity for 3000 observations, and an increase in capacity for 1000 observations, whenever the current capacity limit is reached.

  \begin{lstlisting}
gp = ElasticGPE(2,                       # two input dimensions
                mean = MeanConst(0.),
                kernel = SEArd([0., 0.], 5.),
                logNoise = 0.,
                capacity = 3000,
                stepsize = 1000)
  \end{lstlisting}
  \begin{lstlisting}
GP Exact object:
Dim = 2
Number of observations = 0
Mean function:
Type: MeanConst, Params: [0.0]
Kernel:
Type: SEArd{Float64}, Params: [-0.0, -0.0, 5.0]
  No observation data
  \end{lstlisting}
  \begin{lstlisting}
append!(gp, [1., 2.], 0.4)      # append observation x = [1., 2.] and y = 0.4  
  \end{lstlisting}
  \begin{lstlisting}
GP Exact object:
Dim = 2
Number of observations = 2
Mean function:
Type: MeanConst, Params: [0.0]
Kernel:
Type: SEArd{Float64}, Params: [-0.0, -0.0, 5.0]
Input observations = 
[1.0 1.0; 2.0 2.0]
Output observations = [0.4, 0.4]
Variance of observation noise = 1.0
Marginal Log-Likelihood = -7.184
  \end{lstlisting}

Under the hood, elastic GPs allocates memory for the number of observations specified with the keyword argument \code{capacity} and uses \code{view}s to select only the part of memory that is already filled with actual observations.
Whenever the current capacity \code{c} is reached, memory for \code{c + stepsize} observations is allocated and the old data copied over.
Elastic GPs uses efficient rank-one updates of the Cholesky decomposition that holds the covariance data of the GP.

\begin{figure}[h]
  \centering
  \includegraphics{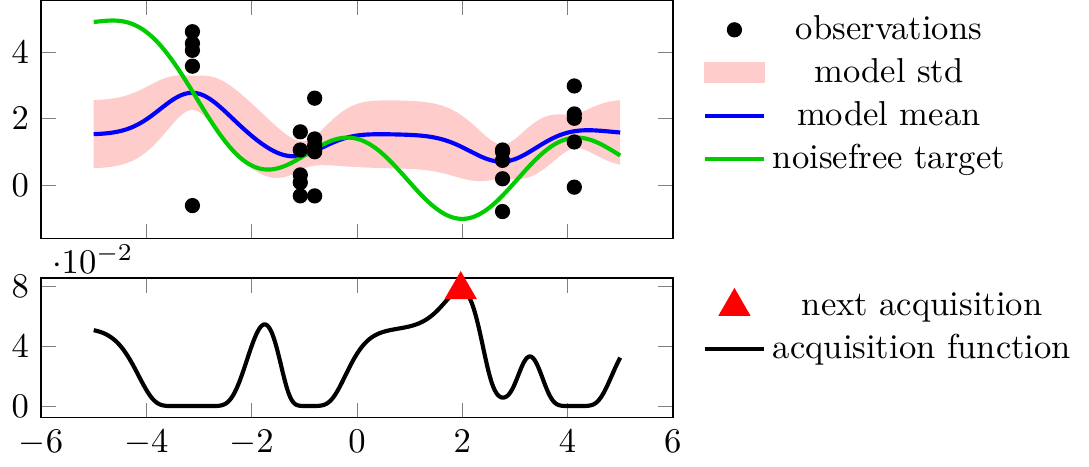}
  \caption{After evaluating the noisy function $f(x)= 0.1\cdot(x - 2)^2 + \cos(\pi/2\cdot x) + \epsilon$, where $\epsilon \sim \mathcal{N}(0,1)$, at five positions five times (black dots) and fitting a Gaussian process (blue line with red standard deviations), the expected improvement acquisition function (black line) peaks near input value 2, where the next acquisition will occur.}
  \label{fig:bayes_opt}
\end{figure}

In the following example we use Bayesian optimization on a not so costly, but noisy one-dimensional objective function, $f(x)= 0.1\cdot(x - 2)^2 + \cos(\pi/2\cdot x) + \epsilon$, where $\epsilon \sim \mathcal{N}(0,1)$, which is illustrated in Figure \ref{fig:bayes_opt}.

  \begin{lstlisting}
using BayesianOptimization, GaussianProcesses
    
f(x) = 0.1*(x[] - 2)^2 + cos(pi/2*x[]) + randn()  # noisy function to minimize

# Choose as a model an elastic GP with input dimensions 1.
 model = ElasticGPE(1, mean = MeanConst(0.), kernel = SEArd([0.], 5.), logNoise = 0.)

 # Optimize the hyperparameters of the GP using maximum likelihood (ML) 
 # estimates every 50 steps woth bounds on the logNoise and
 # bounds for the two parameters GaussianProcesses.get_param_names(model.kernel)
modeloptimizer = MAPGPOptimizer(every = 50,
                                noisebounds = [-2., 3],          
                                kernbounds = [[-1, 0], [4, 10]], 
                                maxeval = 40)
    
opt = BOpt(f, model,
            ExpectedImprovement(),          # type of acquisition function
            modeloptimizer,
            [-5.], [5.],                    # lower- and upperbounds
            repetitions = 5,                # evaluate the function 5 times
            maxiterations = 200,            # evaluate at 200 input positions
            sense = Min,                    # minimize the objective function
            acquisitionoptions = (maxeval = 4000, restarts = 50),
            verbosity = Silent)    
result = boptimize!(opt)
  \end{lstlisting}
  \begin{lstlisting}
  (observerd_optimum = -3.561010774347263, observed_optimizer = [1.8632],
   model_optimum = -1.0408708431654201, model_optimizer = [1.99274])        
  \end{lstlisting}

\pkg{BayesianOptimization.jl} uses automatic differentiation tools in \pkg{ForwardDiff.jl} \citep{RevelsLubinPapamarkou2016} to compute gradients of the acquisition function (\code{ExpectedImprovement()} in the example above).
After evaluating the function at 200 positions, the global minimum of the Gaussian process at \code{model\_optimizer = [1.99274]} is close to the global minimum of the noise-free objective function.

\subsection{Sparse inputs}
In this section we will demonstrate how each of the sparse approximations detailed in Section \ref{sec:sparsegp} can be used. The performance of each sparse method will be demonstrated by fitting a sparse Gaussian process to a set of $n=5000$ data points that are simulated from $f(x)=|x-5|\cos(2x)$,

  \begin{lstlisting}
# The true function we will be simulating from is,
function fstar(x::Float64)
    return abs(x-5)*cos(2*x)
end

$\sigma$y = 10.0           # observation noise
n=5000             # number of observations

Random.seed!(1) # for reproducibility
Xdistr = Beta(7,7)
$\epsilon$distr = Normal(0,$\sigma$y)
x = rand(Xdistr, n)*10
X = Matrix(x')
Y = fstar.(x) .+ rand($\epsilon$distr,n)    
  \end{lstlisting}

The set of inducing point locations $X_{\bu}$ used here will be consistent for each method and are defined explicitly.

  \begin{lstlisting}
Xu = Matrix(quantile(x, [0.2, 0.25, 0.3, 0.35, 0.4, 0.45, 0.5, 0.55, 0.6,
                         0.65, 0.7, 0.98])')    
  \end{lstlisting}

With the inducing point locations defined, we can now fit each of the sparse Gaussian process approximations. The practitioner is free to select from any of the sparse approaches outlined in Section \ref{sec:sparsegp}, each of which can be invoked using the below syntax. The only syntactic deviation is the full scale approach, which requires the practitioner to choose the covariance matrix's local blocks. In this example, $m$ blocks have been created, with a one-to-one mapping between block and inducing point locations.

  \begin{lstlisting}
# Subset of Regressors
gp_SOR = GaussianProcesses.SoR(X, Xu, Y, MeanConst(mean(Y)), k, log($\sigma$y));

# Determinetal Training Conditional
gp_DTC = GaussianProcesses.DTC(X, Xu, Y, MeanConst(mean(Y)), k, log($\sigma$y));

# Fully Independent Training Conditional 
gp_FITC = GaussianProcesses.FITC(X, Xu, Y, MeanConst(mean(Y)), k, log($\sigma$y));

# Full Scale 
inearest = [argmin(abs.(xi.-Xu[1,:])) for xi in x]
blockindices = [findall(isequal(i), inearest) for i in 1:size(Xu,2)]

GaussianProcesses.FSA(X, Xu, blockindices, Y, MeanConst(mean(Y)), k, log($\sigma$y));    
  \end{lstlisting}

Prediction is handled in the same way as a regular Gaussian process, using the \code{predict\_f} function.

\begin{figure}
\begin{tabular}{cc}
  \includegraphics[width=0.49\textwidth]{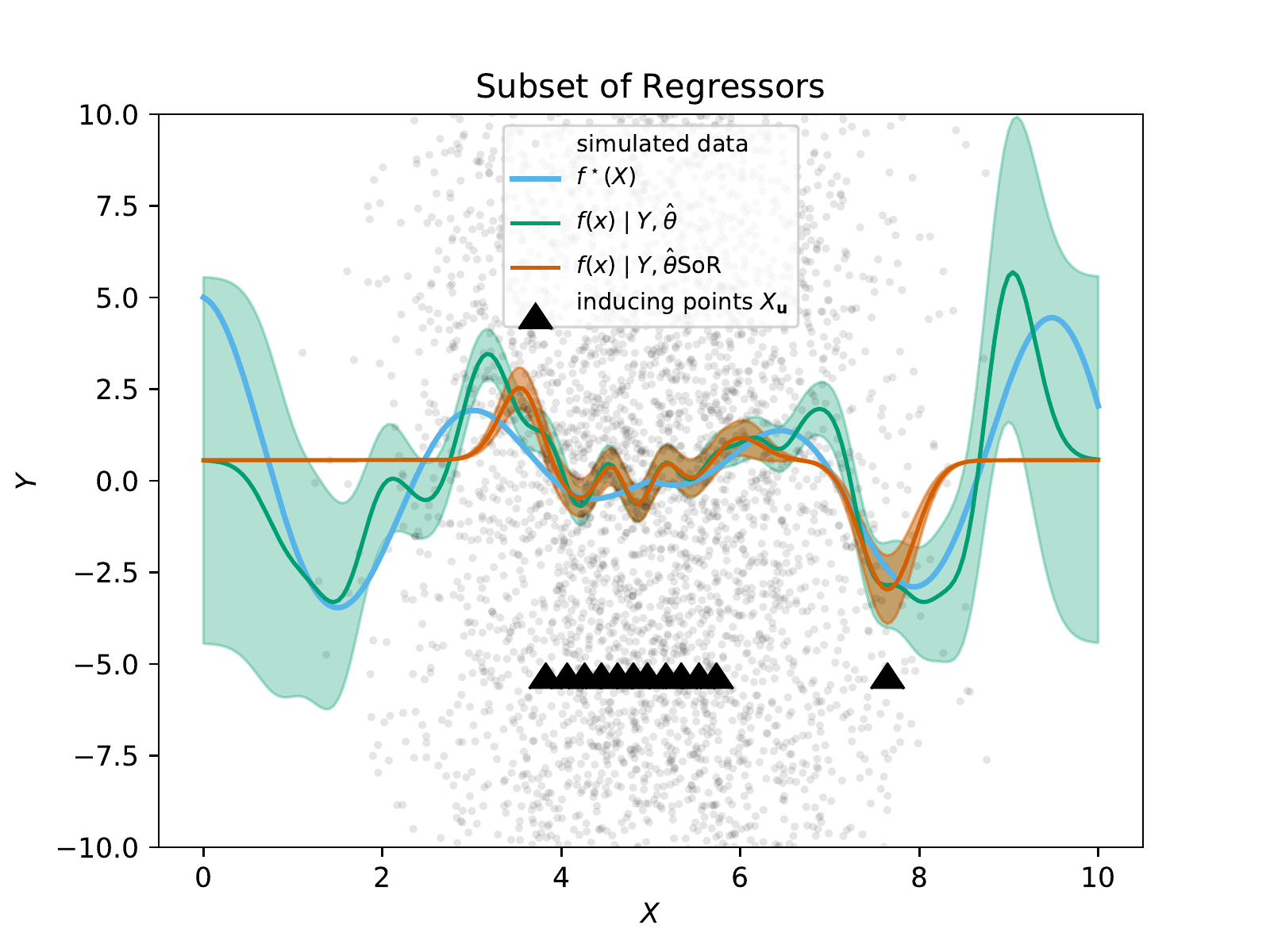} &   \includegraphics[width=0.49\textwidth]{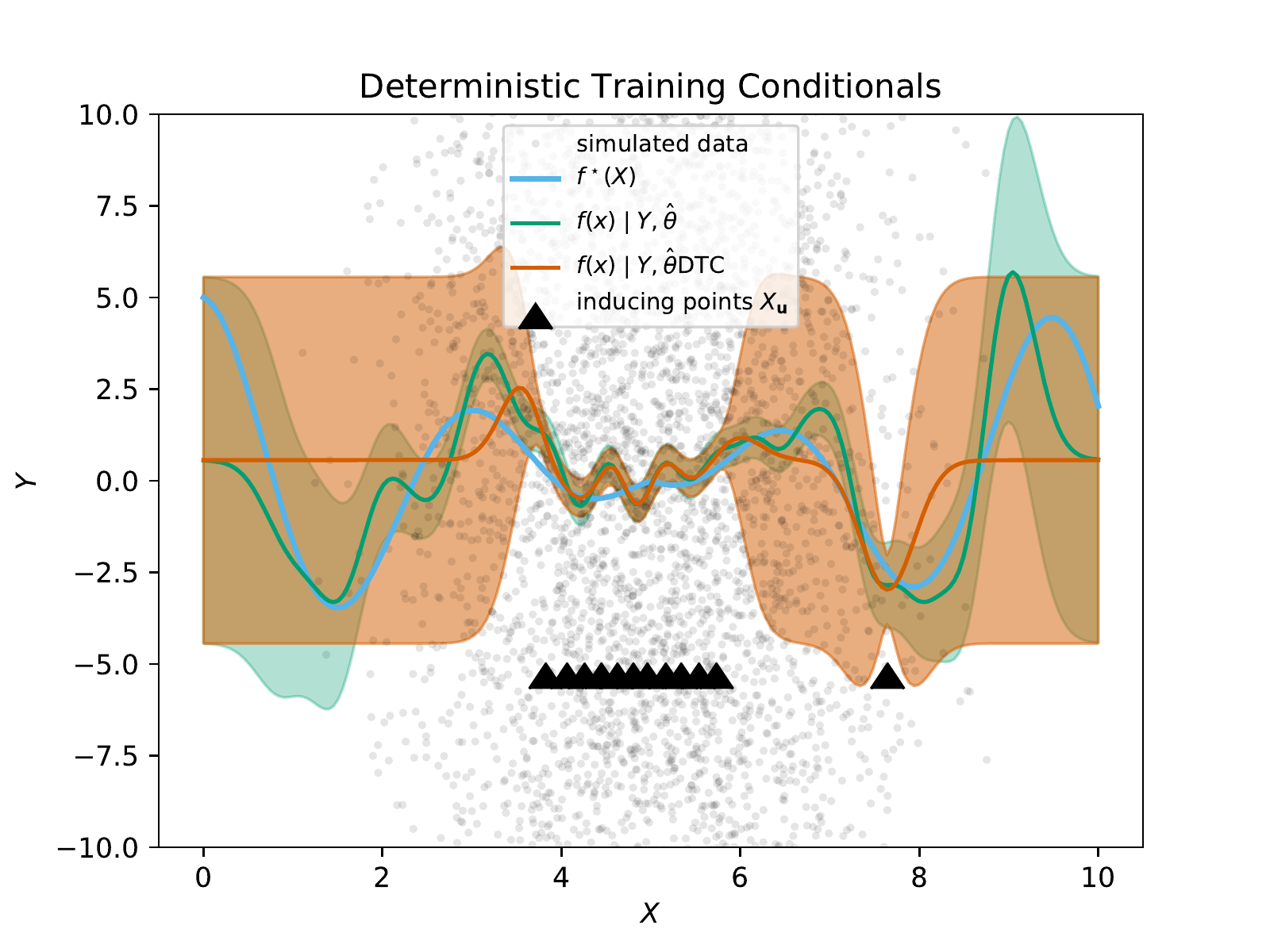} \\
(a) SoR & (b) DTC \\[1pt]
 \includegraphics[width=0.49\textwidth]{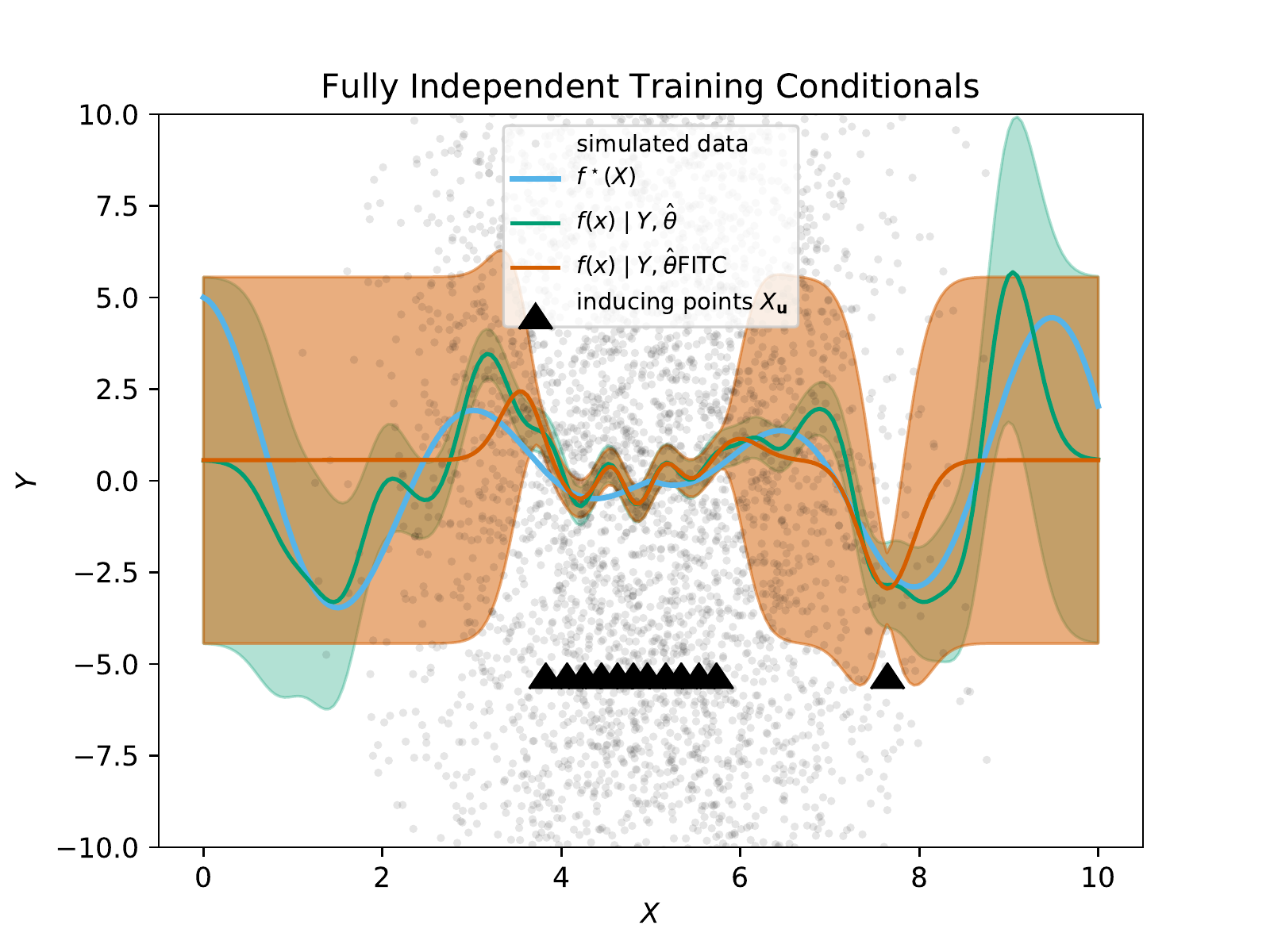} &   \includegraphics[width=0.49\textwidth]{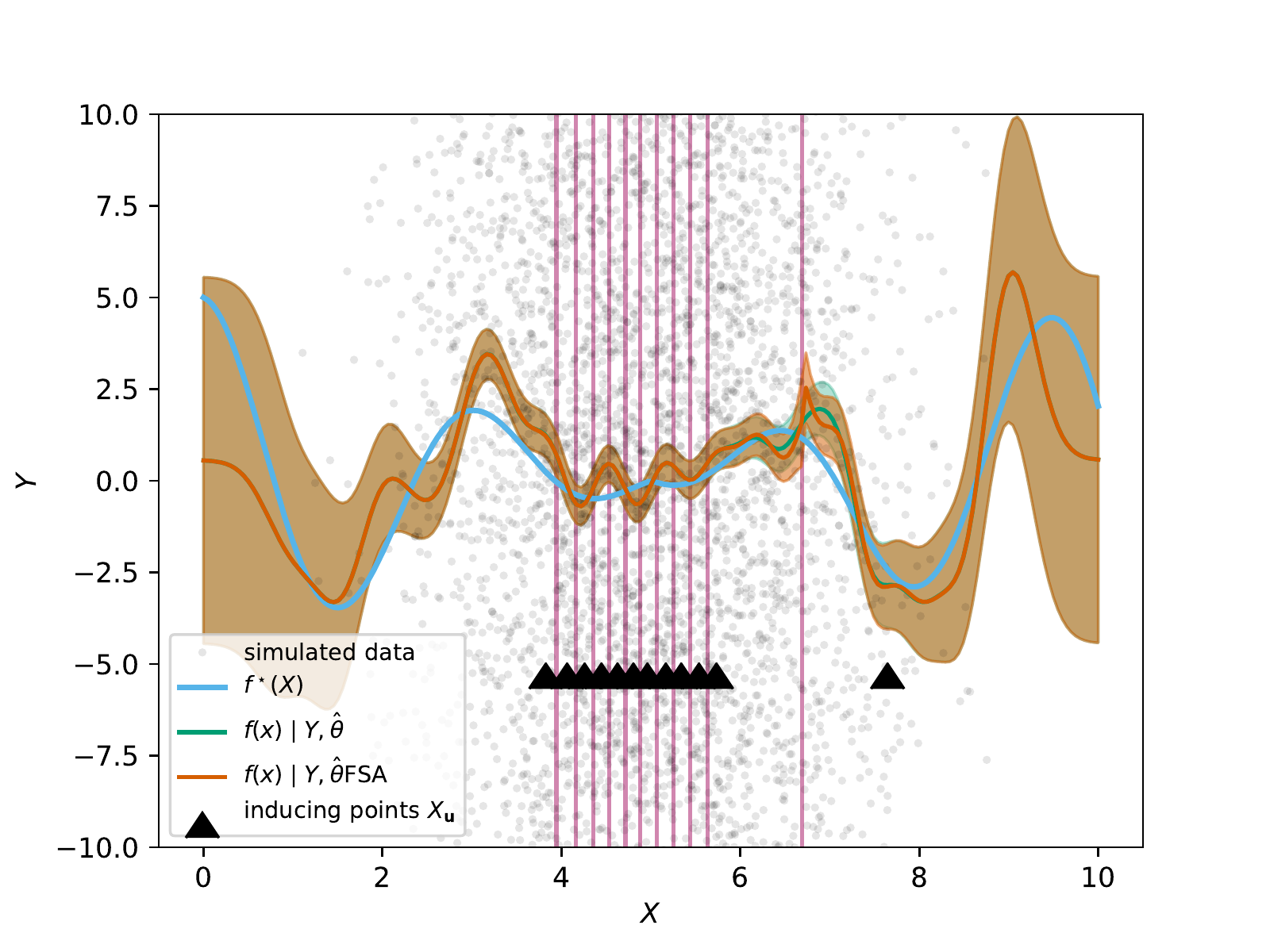} \\
(c) FITC & (d) Full Scale \\
\end{tabular}
\caption{Comparison of sparse approximations to an exact Gaussian process. The vertical purple lines in panel d indicate the dividing lines between blocks where an information discontinuity will occur.}
\label{fig:sparsegps}
\end{figure}

As discussed in Section \ref{sec:sparsegp}, each sparse method yields a computational acceleration, however, this often comes at the cost of poorer predictive inference, as shown in Figure \ref{fig:sparsegps}. This is no more apparent than in the SoR approach, where the posterior predictions are excessively confident, particularly beyond the range of the inducing points. Both the DTC and FITC are more conservative in the predictive uncertainty as the process moves away from the inducing points' location, while sacrificing little in terms of computational efficiency\footnote{All simulations run on a Linux machine with a 1.60GHz i5-8250U CPU and 16GB RAM.} - see Table \ref{tab:sparsegp} for computational timing results.

\begin{table}[tbp]
\centering
\begin{tabular}{lrr}
\hline
           & CPU Runtime (seconds)       & Memory Allocation (MiB) \\ \hline
Exact      & 1.417324          & 572.320                 \\
SoR        & 0.004076          & \textbf{2.032}          \\
DTC        & \textbf{0.003104} & 2.033                   \\
FITC       & 0.022644          & 3.902                   \\
Full Scale & 0.383900          & 156.025                 \\ \hline
\end{tabular}
\caption{The computational efficiency of fitting each of sparse approximations to the training data. }
\label{tab:sparsegp}
\end{table}

\section{Comparison to other packages}
\label{sec:comp-other-pack}

In this section we compare the performance of \pkg{GaussianProcesses.jl} to two leading Gaussian process inference packages for the fundamental task of computing the log-likelihood, and its gradient, in a simulated problem with a Gaussian likelihood.
We use version 4.1 of the \proglang{MATLAB} package \pkg{GPML} \citep{rasmussen2010gaussian,RasmussenNickisch2017}, which was originally written to demonstrate the algorithms in \cite{rasmussen}, and has since become a mature package, often integrating new algorithms from the latest research on Gaussian processes.
The package is mostly written in pure \proglang{MATLAB}, except for a small number of optimisation and linear algebra routines implemented in \proglang{C}.
We also compare to version 1.9.2 of \pkg{GPy} \citep{gpy2014}, a \proglang{python} package dedicated to Gaussian processes, with core components written in \proglang{cython}.
We first simulated $n=3,000$ standard normal observations, each with 10 covariates also simulated as standard normals.
We reuse the same simulated dataset for every benchmark.
In each package, we benchmark the function that updates the log-likelihood and its gradient given a set of parameters, by running it 10 times and reporting the duration of the shortest run.
We compare the packages' performance for a variety of covariance kernels, with all variance, length-scale, or shape parameters set to 1.0.
The results are presented in Table \ref{table:benchmarks}, and the benchmark code for each package is available with the  \pkg{GaussianProcesses.jl} source code.

We find that \pkg{GaussianProcesses.jl} is highly competitive with \pkg{GPML} and \pkg{GPy}.
It has the fastest run-times for all of the 10 kernels considered, including the additive and product kernels.
    
\begin{table}[tbp]
\centering
\begin{tabular}{lrrr}
\hline
Kernel                                 & \pkg{GaussianProcesses.jl} & \pkg{GPy}  & \pkg{GPML}  \\ 
\hline
\code{fix(SE(0.0,0.0),} $\sigma)$      & \textbf{730}     & 1255     &       \\
\code{SE(0.0,0.0)}                     & \textbf{800}     & 1225     & 1131  \\
\code{Matern(1/2,0.0,0.0)}             & \textbf{836}     & 1254     & 1246  \\
\code{Masked(SE(0.0,0.0), [1]))}       & \textbf{819}     & 1327     & 1075  \\
\code{RQ(0.0,0.0,0.0)}                 & \textbf{1252}    & 1845     & 1292  \\
\code{SE(0.0,0.0) + RQ(0.0,0.0,0.0)}   & \textbf{1351}    & 1937     & 1679  \\
\code{Masked(SE(0.0,0.0), [1])} \\ \code{ +  Masked(RQ(0.0,0.0,0.0), collect(2:10))} & \textbf{1562}    & 1893     & 1659  \\
\code{(SE(0.0,0.0) + SE(0.5,0.5)) * RQ(0.0,0.0,0.0)}  & \textbf{1682}    & 1953     & 2127  \\
\code{SE(0.0,0.0) * RQ(0.0,0.0,0.0)}    & \textbf{1614}    & 1929     & 1779  \\
\code{(SE(0.0,0.0) + SE(0.5,0.5)) * RQ(0.0,0.0,0.0)}  & \textbf{1977}    & 2042     & 2206  \\
\hline
\end{tabular}
\caption{
    Benchmark results, ordered by running time in \pkg{GaussianProcesses.jl}.
    All times are in milliseconds, and the fastest run-time is bolded.
    The kernels are labelled with their function name from the package: 
    \code{SE} is a squared exponential kernels;
    \code{RQ} is a rational quadratic kernel;
    \code{Matern(1/2,..)} is a Mat\'ern 1/2 kernel;
    sum and product kernels are indicated with \code{+} and \code{*};
    \code{fix(SE(0.0,0.0), $\sigma$)} has a fixed variance parameter (not included in the gradient);
    and \code{Masked(k,[dims])} means the \code{k} kernel is only applied to the covariates \code{dims}.
    \label{table:benchmarks}
}
\end{table}

\section{Future developments}
\label{sec:future-developments}

\pkg{GaussianProcesses.jl} is a fully formed package providing a range of kernel, mean and likelihood functions, and inference tools for Gaussian process modelling with Gaussian and non-Gaussian data types. The inclusion of new features in the package is ongoing and the development of the package can be followed via the Github page\footnote{https://github.com/STOR-i/GaussianProcesses.jl}. The following are package enhancements currently under development:
\begin{itemize}
\item Variational approximations - Currently the package uses MCMC for inference with non-Gaussian likelihoods. MCMC has good theoretical convergence properties, but can be slow for large data sets. Variational approximations \cite{opper2009variational} using, for example, mean-field, have been widely used in the Gaussian process literature, and while not exact, can produce highly accurate posterior approximations.
\item Automatic differentiation - The package provides functionality for maximum likelihood estimation of the hyperparameters, or Bayesian inference using an MCMC algorithm. In both cases, these functions require gradients to be calculated for optimisation or sampling. Currently, derivatives of functions of interest (e.g. log-likelihood function) are hand-coded for computational efficiency. However, recent tests have shown that calculating these gradients using automatic differentiation does not incur a significant additional computational overhead. In the future, the package will move towards implementing all gradient calculations using automatic differentiation. The main advantage of this approach is that users will be able to add new functionality to the package more easily, for example creating a new kernel functions.
\item Gaussian Process Latent Variable Model (GPLVM) - Currently the package is well suited for supervised learning tasks, whereby an observational value exists for each input. GPLVMs are a probabilistic dimensionality reduction method that use Gaussian processes to learn a mapping between an observed, possibly very high-dimensional, variable and a reduced dimension latent space. GPLVMs are a popular method for dimensionality reduction as they transcend principal component analysis by learning a non-linear relationship between the observations and corresponding latent space. Furthermore, a GPLVM is also able to express the uncertainty surrounding the structure of the latent space. In the future, the package will support the original GPLVM of \cite{lawrence2004gaussian}, and its Bayesian counterpart \cite{titsias2010bayesian}.
\end{itemize}

\section{Acknowledgments}
\label{sec:acknowledgements}

The authors would like to thank the users of the \pkg{GaussianProcesses.jl} package who have helped to support its development. CN gratefully acknowledges the support of EPSRC grants EP/S00159X/1 and EP/R01860X/1.
JB was supported by Swiss National Science Foundation (no.  200020\_184615) and by the European Union Horizon 2020 Framework Program under grant agreement no. 785907 (HumanBrain Project, SGA2).
TP is supported by the Data Science for the Natural Environment project (EPSRC grant number EP/R01860X/1).

\bibliographystyle{apalike}
\bibliography{library,library1}

\bibliography{refs}

%% -- Appendix (if any) --------------------------------------------------------
%% - After the bibliography with page break.
%% - With proper section titles and _not_ just "Appendix".

\newpage

\begin{appendix}

  \section{Available functions}
\label{sec:available-functions}

\begin{sidewaystable}
  \centering
  \begin{tabular}{|l|l|l|l|}
    \hline
    Function & Description & $k_\theta(\bx,\bx^*)=$ &$\btheta$ \\
    \hline\hline
    \code{Const}  & Constant  & $\sigma^2$ & $\log \sigma$  \\
    \hline
     \code{Lin}   & Linear ARD & $\bx^\top L^{-2} \bx^*, \quad L = \mbox{diag}(\ell_1,\ldots,\ell_d)$ & $(\log \ell_1, \ldots, \log \ell_d)$  \\
    \hline
      \code{Lin}       & Linear Iso & $\bx^\top \bx^*/\ell^2$& $\log \ell$  \\
    \hline
    \shortstack{\code{Matern(1/2,...)} \\ \code{Matern(3/2,...)}  \\ \code{Matern(5/2,...)} }         & \shortstack{Mat\'ern ARD (1/2) \\ Mat\'ern ARD (3/2) \\ Mat\'ern ARD (5/2)} &  \shortstack{$\sigma^2 \exp(-|\bx-\bx^*|/L), \quad L=\mbox{diag}(\ell_1,\ldots,\ell_d)$ \\ $\sigma^2(1+\sqrt{3}|\bx-\bx^*|/L) \exp(-\sqrt{3}|\bx-\bx^*|/L)$ \\ $\sigma^2(1+\sqrt{5}|\bx-\bx^*|/L + 5|\bx-\bx^*|^2/3L^2) \exp(-\sqrt{5}|\bx-\bx^*|/L)$} & \shortstack{$(\log \ell_1, \ldots, \log \ell_d, \log \sigma)$ \\ $(\log \ell_1,  \ldots, \log \ell_d, \log \sigma)$ \\ $(\log \ell_1,  \ldots, \log \ell_d,\log \sigma)$} \\
    \hline
         \shortstack{\code{Matern(1/2,...)} \\ \code{Matern(3/2,...)}  \\ \code{Matern(5/2,...)} }   & \shortstack{Mat\'ern Iso (1/2) \\ Mat\'ern Iso (3/2) \\ Mat\'ern Iso (5/2)} & \shortstack{$\sigma^2 \exp(-|\bx-\bx^*|/\ell)$  \\ $\sigma^2(1+\sqrt{3}|\bx-\bx^*|/\ell) \exp(-\sqrt{3}|\bx-\bx^*|/\ell)$ \\ $\sigma^2(1+\sqrt{5}|\bx-\bx^*|/\ell + 5|\bx-\bx^*|^2/3\ell^2) \exp(-\sqrt{5}|\bx-\bx^*|/\ell)$} & \shortstack{$(\log \ell, \log \sigma)$ \\ $(\log \ell, \log \sigma)$ \\ $(\log \ell, \log \sigma)$} \\
    \hline
     \code{SE}        & Squared exponential ARD & $\sigma^2 \exp(-(\bx-\bx^*)^\top L^{-2}(\bx-\bx^*)/2)$ \quad $L=\mbox{diag}(\ell_1,\ldots,\ell_d)$ & $(\log \ell_1,\ldots, \log \ell_d,\log \sigma)$ \\
    \hline
             & Squared exponential Iso & $\sigma^2 \exp(-(\bx-\bx^*)^\top (\bx-\bx^*)/2\ell^2)$ & $(\log \ell, \log \sigma)$  \\
    \hline
     \code{Periodic}        & Periodic  & $\sigma^2 \exp(-2 \sin^2(\pi|\bx-\bx^*|/p)/\ell^2)$ & $(\log \ell, \log \sigma, \log p)$ \\
    \hline
     \code{Poly}  & \shortstack{Polynomial \\ (degree (d) is user defined)} & $\sigma^2(\bx^\top \bx^* + c)^d$ & $(\log c, \log \sigma)$  \\
    \hline
     \code{Noise}   & Noise  & $\sigma^2 \delta(\bx-\bx^*)$ & $(\log \sigma)$ \\
    \hline
    \code{RQ}    & Rational Quadratic ARD & $\sigma^2(1+(\bx-\bx^*)^\top L^{-2} (\bx-\bx^*)/2\alpha)^{-\alpha}$ \quad $L=\mbox{diag}(\ell_1,\ldots,\ell_d)$& $(\log \ell_1, \ldots, \log \ell_d, \log \sigma, \log \alpha)$  \\
    \hline
     \code{RQ}   & Rational Quadratic Iso & $\sigma^2(1+(\bx-\bx^*)^\top (\bx-\bx^*)/2\alpha\ell^2)^{-\alpha}$ & $(\log \ell, \log \sigma, \log \alpha)$  \\
    \hline
    \code{FixedKernel}  & \shortstack{Fixed kernels \\ (fix some hyperparameters)} & $k_\theta(\bx,\bx^*)$ & $\theta^\prime \subseteq \theta$  \\
    \hline
    \code{Masked}  & \shortstack{Masked kernels \\ only active dimensions \\ $i \subseteq \{1,,\ldots,d\}$} & $k_\theta(\bx_i,\bx_i^*)$ & $\emptyset$  \\
    \hline
    \code{*}  & Product kernels & $\prod_i k_\theta(\bx,\bx^*)$ & $\emptyset$   \\
    \hline
     \code{+}   & Sum kernels & $\sum_i k_\theta(\bx,\bx^*)$ & $\emptyset$  \\
    \hline
  \end{tabular}
  \caption{Table of available kernel functions. ARD: Automatic Relevance Determination, Iso: Isotropic}
\label{tab:kernels}
\end{sidewaystable}

\begin{sidewaystable}
  \centering
  \begin{tabular}{|l|l|l|l|l|}
    \hline
    Function & Description & $p(y_i\,\given \,f_i,\btheta)=$ & Transform &$\btheta=$ \\
    \hline\hline
    \code{BernLik}  & Bernoulli - $y_i\in\{0,1\}$ & $g_i^{y_i}(1-g_i)^{(1-{y_i})}$  & $g_i=\Phi(f_i)$ & $f_i$ \\
    \hline
     \code{BinLik}  & Binomial - $y_i\in\{0,1,\ldots,n\}$ & $\frac{y_i!}{n!(n-y_i)!} g_i^{y_i}(1-g_i)^{(1-{y_i})}$  & $g_i=\frac{\exp(f_i)}{1+\exp(f_i)}$ & $f_i$\\
    \hline
    \code{ExpLik}  & Exponential - $y_i \in \mathbb{R}_+$ & $g_i\exp(-g_iy_i)$ & $g_i = \exp(-f_i)$ & $f_i$ \\
    \hline
     \code{GaussLik}   & Gaussian - $y_i \in \mathbb{R}$& $1/\sqrt{2\pi\sigma^2} \exp{(-(y_i-f_i)^2/2\sigma^2)}$  &$f_i$ & $(f_i,\log \sigma)$\\
    \hline
     \code{PoisLik} & Poisson - $y_i \in \mathbb{N}_0$ & $g_i^{y_i}\exp(-g_i)/y_i!$  & $g_i=\exp(f_i)$ & $f_i$\\
     \hline
     \code{StuTLik}  & Student-t - $y_i \in \mathbb{R}$ & $\frac{\Gamma((\nu+1)/2)}{\sqrt{\Gamma(\nu/2)\pi\nu}\sigma}(1+\frac{1}{\nu}(\frac{(y_i-f_i)}{\sigma})^2)^{-(\nu+1)/2}$ &$f_i$ &$(f_i,\log \sigma)$ \\
     \hline
  \end{tabular}
  \caption{List of available likelihood functions}
  \label{tab:likelihoods}

  \bigskip\bigskip
  
  \centering
  \begin{tabular}{|l|l|l|l|}
    \hline
    Function & Description & $m_\theta(\bx)=$ &$\btheta=$ \\
    \hline\hline
    \code{MeanZero}  & Zero & $0$ & $\emptyset$  \\
        \hline
    \code{MeanConst}  & Constant   & $\btheta$, \quad $\btheta=(\theta_1,\ldots,\theta_d)$ & $\btheta$  \\
    \hline
        \code{MeanLin}  & Linear & $\bx^\top \btheta$, \quad $\btheta=(\theta_1,\ldots,\theta_d)$  & $\btheta$  \\
    \hline
        \code{MeanPoly}  & Polynomial  (of degree $D$) & $\sum_{j=1}^D \btheta_j\bx^j$, \quad $\btheta_j = (\theta_{1j},\ldots,\theta_{dj})$  & $\btheta_j$ \quad $\forall j \in \{1,2,\ldots,D\}$  \\
    \hline
        \code{+}  & Sum  & $\sum_i m_\theta(\bx)$ & $\emptyset$ \\
    \hline
        \code{*}  & Product &  $\prod_i m_\theta(\bx)$ & $\emptyset$ \\
    \hline
  \end{tabular}
  \caption{List of available mean functions}
\label{tab:means}

  \bigskip\bigskip

  \centering
  \begin{tabular}{|l|l|l|l|}
    \hline
    Function & Description & $q(\bff|\bu)=$ &$q(\bff^*|\bu)=$ \\
    \hline\hline
    \code{SoR}  & Subset of regressors & $\mathcal{N}(\Kfu \Kuu^{-1}\bu, \mathbf{0})$ & $\mathcal{N}(\Kxu \Kuu^{-1}\bu, \mathbf{0})$  \\
        \hline
    \code{DTC}  & Deterministic Training Conditional & $\mathcal{N}(\Kfu \Kuu^{-1}\bu, \mathbf{0})$ & $\mathcal{N}(\Kxu \Kuu^{-1}\bu, \Kxx-\Qxx)$  \\
    \hline
        \code{FITC}  & Fully Independent Training Conditional & $\mathcal{N}(\Kfu \Kuu^{-1}\bu, \text{diag}\left[\Kff - \Qff\right])$  & $\mathcal{N}(\Kxu \Kuu^{-1}\bu, \Kxx-\Qxx)$ \\
    \hline
        \code{FSA}  & Full Scale Approximation & $\mathcal{N}(\Kfu \Kuu^{-1}\bu, \text{blockdiag}\left[\Kff - \Qff\right])$   & $\mathcal{N}(\Kxu \Kuu^{-1}\bu, \Kxx-\Qxx)$  \\
    \hline
  \end{tabular}
  \caption{List of available sparse approximations and their imposed conditional distributions}
\label{tab:sparse}

\end{sidewaystable}

\end{appendix}

%% -----------------------------------------------------------------------------

\end{document}